\documentclass[twocolumn]{aastex6}
\RequirePackage{lineno}
 \usepackage{amsmath}
 \usepackage{amssymb}
 \usepackage{amsthm}
 \usepackage{natbib,aasdefs,url,bm}
 \usepackage{array}
 \usepackage{float}
 \usepackage{graphicx}
 \usepackage{subfigure}
 \usepackage{color}

\newcommand{\pd}{\partial}  
\newcounter{ichi}
\newcounter{ni}
\newcounter{san}
\newcounter{yon}
\slugcomment{}

\shorttitle{High-Energy Emission from Interacting Supernovae}
\shortauthors{Murase et al.}

\begin{document}

\title{High-Energy Emission from Interacting Supernovae:\\
New Constraints on Cosmic-Ray Acceleration in Dense Circumstellar Environments}
\author{
Kohta Murase\altaffilmark{1,2,3,4}, Anna Franckowiak\altaffilmark{5}, Keiichi Maeda\altaffilmark{6}, Raffaella Margutti\altaffilmark{7}, John F. Beacom\altaffilmark{8,9,10}}
\altaffiltext{1}{Department of Physics, Pennsylvania State University, University Park, Pennsylvania 16802, USA}
\altaffiltext{2}{Department of Astronomy \& Astrophysics, Pennsylvania State University, University Park, Pennsylvania 16802, USA}
\altaffiltext{3}{Center for Particle and Gravitational Astrophysics, Pennsylvania State University, University Park, Pennsylvania 16802, USA}
\altaffiltext{4}{Center for Gravitational Physics, Yukawa Institute for Theoretical Physics, Kyoto University, Kyoto, Kyoto 606-8502, Japan}
\altaffiltext{5}{Deutsches Elektronen-Synchrotron (DESY), Platanenallee 6, Zeuthen D-15738, Germany}
\altaffiltext{6}{Department of Astronomy, Kyoto University, Kyoto, Kyoto 606-8502, Japan}
\altaffiltext{7}{Department of Physics and Astronomy, Northwestern University, Evanston, IL 60208, USA}
\altaffiltext{8}{Center for Cosmology and Astro Particle Physics, Ohio State University, Columbus, Ohio 43210, USA}
\altaffiltext{9}{Department of Physics Ohio State University, Columbus, Ohio 43210, USA}
\altaffiltext{10}{Department of Astronomy, Ohio State University, Columbus, Ohio 43210, USA}


\begin{abstract}
Supernovae (SNe) with strong interactions with circumstellar material (CSM) are promising candidate sources of high-energy neutrinos and gamma rays, and have been suggested as an important contributor to Galactic cosmic rays beyond ${10}^{15}$~eV. Taking into account the shock dissipation by a fast velocity component of SN ejecta, we present comprehensive calculations of the non-thermal emission from SNe powered by shock interactions with a dense wind or CSM. Remarkably, we consider electromagnetic cascades in the radiation zone and subsequent attenuation in the pre-shock CSM. A new time-dependent phenomenological prescription provided by this work enables us to calculate gamma-ray, hard X-ray, radio, and neutrino signals, which originate from cosmic rays accelerated by the diffusive shock acceleration mechanism.  
We apply our results to SN IIn 2010jl and SN Ib/IIn 2014C, for which the model parameters can be determined from the multi-wavelength data.  
For SN 2010jl, the more promising case, by using the the latest {\it Fermi} Large Area Telescope (LAT) Pass 8 data release, we derive new constraints on the cosmic-ray energy fraction, $\epsilon_{p}\lesssim0.05-0.1$. We also find that the late-time radio data of these interacting SNe are consistent with our model. 
Further multi-messenger and multi-wavelength observations of nearby interacting SNe should give us new insights into the diffusive shock acceleration in dense environments as well as pre-SN mass-loss mechanisms. 
\end{abstract}



\section{Introduction}
Supernovae (SNe) have been widely believed to be the main contributors to Galactic cosmic rays (CRs). 
The observed CR flux can be explained if $\sim10$\% of the SN ejecta energy, ${\mathcal E}_{\rm ej}\sim10^{51}$~erg, is converted into the kinetic energy of accelerated CR ions. It has been believed that the diffusive shock acceleration (DSA) mechanism is responsible for particle acceleration in supernova remnants (SNRs)~\citep[see, e.g.,][for a review]{Drury:1983zz}, and SNRs have been observed in gamma rays with gamma-ray observatories such as {\it Fermi} and imaging atmospheric Cherenkov telescopes (IACTs) such as H.E.S.S., MAGIC, and VERITAS. Recent detailed observations at sub-GeV energies have led to the discovery of a low-energy cutoff due to neutral pion decay, which is regarded as evidence of hadronic gamma-ray emission from $\pi^0\rightarrow2\gamma$~\citep[see][and references therein]{Funk:2015ena}. Independent of these phenomenological and observational arguments, state-of-art particle-in-cell simulations have also revealed that CR ions carry $\sim10$\% of the SN ejecta energy via the DSA mechanism~\citep[e.g.,][]{Caprioli:2015cda}. 

How early is high-energy non-thermal emission anticipated? Obviously, most of the SN explosion energy remains as bulk kinetic energy until the ejecta begins to decelerate at the Sedov radius. Thus, it has been widely believed that a negligible energy fraction of the SN ejecta energy can be used for high-energy emission in the first several days to months after the explosion. The hydrodynamical evolution of SNe is divided into several phases. Just after the core collapse, a SN shock propagates inside a stellar core or envelope, and the shock is radiation mediated or even collisional~\citep{Weaver:1976}. As the shock reaches the surface of the progenitor star, the photon diffusion time becomes short enough, and the shock breakout (at which the SN emission begins to escape) occurs~
\citep[e.g.,][]{Matzner:1998mg}. 
Then the SN shock starts to sweep up circumstellar material (CSM)\footnote{Strictly speaking, the external material can be a dense wind or CSM shell or even an inflated envelope.} and efficient CR acceleration begins to operate as in SNRs. However, for mass-loss rates that are typical of evolved massive stars, $\sim{10}^{-6}-{10}^{-4}~{\rm M}_\odot~{\rm yr}^{-1}$, high-energy neutrino and gamma-ray emission is detectable only for nearby SN events~\citep{Murase:2017pfe}, because the SN ejecta are largely in the free expansion phase.   

Interestingly, an increasing sample of SNe discovered by recent massive searches for optical and near-IR transients has revealed that strong shock interactions between the SN ejecta and a dense wind or CSM shell occur in various classes of explosive transients, as suggested by observations of Type IIn, Ibn, Ia-CSM, and Type II super-luminous supernovae~\citep[e.g.,][]{Smith:2006dk,Miller:2008jy,Ofek:2013mea,Ofek:2013jua,Margutti:2013pfa,Ofek:2013afa,Ofek:2014fua,Margutti:2016wyh}. This suggests that the pre-explosion eruption of the stellar material or inflation of its envelope might be ubiquitous~\citep[see][for a review]{Smith:2014txa}. Although the origin of extended material is not well known, significant mass ($\sim{10}^{-3}-10~{\rm M}_\odot$) may be ejected $\sim0.1-1$~yr before the core-collapse event~\citep[e.g.,][]{Immler:2007mk,Smith:2007cb,Smith:2012nw,Fransson:2013qya,2014MNRAS.442.1166M,Milisavljevic:2015bli}. For such objects, the energy dissipated by the shock is significantly enhanced~\citep[see a review][]{Chevalier:2016hzo}, and a substantial amount of CRs will be produced even in the early phase. 
\cite{Murase:2010cu} and \cite{Murase:2013kda} suggested that non-thermal signatures (including neutrinos, gamma rays, X rays, submm and radio waves) can be used as a probe of the early stage of CR acceleration in SNe IIn. 
The ambient density is so high that most CRs should be converted into neutrino and gamma-ray emission (i.e., the system is ``calorimetric''), in which the CR acceleration can be directly tested by high-energy observations via modeling of optical emission from SNe~\citep{Margutti:2013pfa,TheFermiLAT:2015kla}. 
CR acceleration in dense environments has also been seen by gamma-ray observations of novae~\citep{Ackermann:2014vfa}. 
It is worth mentioning that interacting SNe and Type IIn SNe are much more energetic than novae and they have been suggested as Pevatrons, in particular accelerators of CRs beyond the {\it knee} energy at $10^{15.5}$~eV~\citep{Sveshnikova:2003sa,Murase:2013kda,Zirakashvili:2015mua}. They have also been discussed as one of the origins of IceCube's neutrinos above 0.1~PeV~\citep{Zirakashvili:2015mua,Petropoulou:2017ymv}. Thus, searching for high-energy emission will give us clues to the CR origin as well as mechanisms of early-time particle acceleration and pre-explosion mass losses. 

\begin{figure}[tb]
\centering
\includegraphics[width=\linewidth]{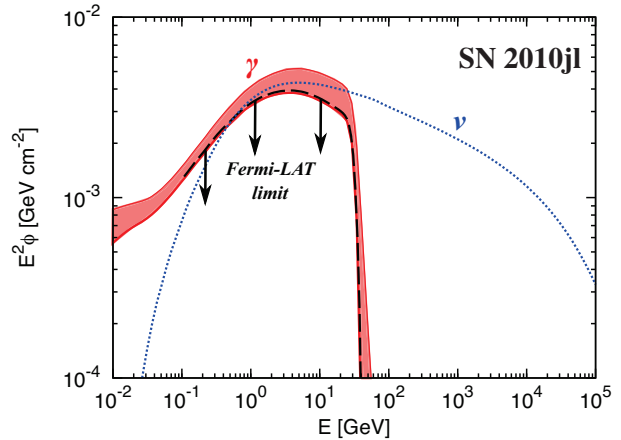}
\caption{Synthetic spectra for hadronic gamma-ray and all-flavor neutrino spectra from SN IIn 2010jl. The CR parameters are $\epsilon_p=0.05~(\epsilon_{\rm sn}/0.25)$ and $s_p=2.2$, and the SN parameters are determined by observations~\citep{Ofek:2013afa}. The upper bound by the one-year {\it Fermi}-LAT observation is indicated by the dashed line with arrows. 
The upper solid curve of the (red) shaded region indicates the prediction for gamma rays from the radiation zone, wheres the lower solid curve corresponds to the case where gamma-ray attenuation in the pre-shock CSM is implemented. The dotted curve represents the predicted neutrino spectrum. 
Note that the fluences integrated from 20~d to 316~d are shown. See text for details.
}
\end{figure}

In this work, we present results of detailed calculations of non-thermal emission from interacting SNe. In the early stages of strongly interacting SNe such as Type IIn SNe, copious thermal photons are present, which inevitably initiate electromagnetic cascades via two-photon annihilation and subsequent synchrotron and inverse-Compton radiation. In addition, gamma rays and radio waves may be attenuated during their propagation in the CSM. The model is described in Section~2, and we apply our phenomenological time-dependent model to SN 2010jl (see Section~3). In particular, using the Pass 8 data of {\it Fermi} Large Area Telescope (LAT), we perform detailed analyses of gamma-ray emission from SN 2010jl and compare the limits with predictions of the theoretical model. To convey one of the main results in this work, we first show predicted neutrino and gamma-ray fluences together with the upper limit by {\it Fermi} LAT in Figure~1. The comparison between the theory and data implies that the energy fraction of CRs is constrained to be $\epsilon_p\lesssim0.05-0.1$. 
In Section~4, we provide a simplified time-integrated model, which can be useful when detailed data are not available. The model is then applied to strongly interacting SN Ib 2014C, in addition to SN 2010jl. In Section~5, we additionally discuss the gamma-ray detectability of nearby SNe, and the results are summarized in Section~6. 
We use the notation $Q={10}^xQ_x$ in CGS units unless noted otherwise.

\section{High-Energy Emission from Interacting Supernovae}
\subsection{Overview}
For a wide range of parameters of SNe indicating interactions with a dense CSM\footnote{We stress that the physical situation is quite different in ordinary SNe such as Type II-P and II-L SNe, in which the CSM is not dense enough that we can naturally expect that the conventional DSA operates~\citep{Murase:2017pfe}. On the other hand, in Type IIn SNe, the shock is initially radiation-mediated, so the onset of CR acceleration is uncertain and worth investigating.}, it has been shown that the SN shock becomes collisionless and non-radiation mediated when the Thomson optical depth $\tau_T$ is less than $\sim c/V_s$~\citep{Murase:2010cu,Katz:2011zx}. 
Note that the formation of collisionless shocks around the photon breakout radius $R_{\rm bo}$ would not occur for a steep density profile expected in the outer stellar envelope. The maximum velocity of the pre-shock CSM is limited by the radiation pressure, which implies that the collisionless shock may form at $R_s\gtrsim{[2(3-w)/(w-1)]}^{1-w}R_{\rm bo}$, where $R_s$ is the shock radius and $w(<3)$ is the ambient density slope\footnote{Radiation from downstream can accelerate the CSM up to $\sim K{\mathcal E}_{\rm rad}/(4\pi R_s^2c)$, where $K$ is the opacity.  In the spherical geometry, the radiation energy ${\mathcal E}_{\rm rad}$ is limited by $\sim M_{\rm cs}(<R_s)V_s^2/2$, where $M_{\rm cs}(<R_s)=\int^{R_s} dR \, 4\pi R^2 \varrho_{\rm cs}$. Noting $\varrho_{\rm cs}(R_{\rm bo})\approx(w-1)c/(KV_sR_{\rm bo})$ from the breakout condition ($\tau_T\sim c/V_s$), we obtain one of the necessary conditions for the onset of collisionless shocks, which implies that CR acceleration occurs for $w\lesssim3$. Whereas high-energy neutrinos are not promising for breakout emission from the stellar envelope that has a steep density profile, efficient neutrino emission is naturally expected for the breakout from a dense, wind-like or shell-like CSM~\citep{Murase:2017pfe}.}. With an CSM (for which $w\sim0-2$ is typically expected), the shock should become collisionless around $R_{\rm bo}$.  Unless the CSM density is too large for Coulomb losses to be relevant~\citep{Murase:2013kda}, the CR acceleration will also begin after photons start to escape at the breakout time, $t_{\rm bo}$.  
Then, CRs interact with CSM nucleons via $pp$ interactions, leading to neutrinos, gamma rays, and X rays, followed by longer timescale radio/submm emission from secondary electrons and positrons. 

How much is the dissipation energy? In the frequently adopted simplified SN ejecta model, the dissipation energy is estimated to be $M_{\rm cs}{\mathcal E}_{\rm ej}/(M_{\rm ej}+M_{\rm cs})$~\citep[e.g.,][]{Murase:2010cu,Petropoulou:2016zar}, where $M_{\rm cs}$ is the CSM mass and $M_{\rm ej}$ is the ejecta mass. However, the realistic SN ejecta has a density structure as a function of velocity, and there is a high-velocity tail. \cite{Murase:2017pfe} considered a time-dependent model, using a self-similar solution \citep{Chevalier:1982,Nadezhin:1985,Moriya:2013hka}. 
In this more detailed model, as a function of the ejecta velocity ($V$), the dissipated energy is given by ${\mathcal E}_{\rm ej}(>V)\propto V^{5-\delta}$, where ${\mathcal E}_{\rm ej}(>V)\equiv\int_V \, dV' (d{\mathcal E}_{\rm ej}/dV')$ and $\delta$ is the index of the outer ejecta density profile introduced via $\varrho_{\rm ej}\propto t^{-3}{(r/t)}^{-\delta}$. 
Note that ${\mathcal E}_{\rm ej}(>V)$ depends on $V$, which is different from the ``total'' ejecta energy ${\mathcal E}_{\rm ej}$.
Here one should keep in mind that the higher-velocity ejecta is more efficiently dissipated, which significantly enhances the detectability of high-energy emission compared to the simple estimate~\citep{Murase:2017pfe}.
While $\delta=10$ (for a convective core with a radiative envelope) and $\delta=12$ (for a radiative core with a convective envelope) are often assumed~\citep{Matzner:1998mg}, realistic values of $\delta$ are uncertain for Type IIn SNe, because the ejecta may already experience some CSM interactions prior to the major dissipation~\citep{vanMarle:2010ak}. Note that smaller values of $\delta$ are also indicated for trans-relativistic SNe that are associated with low-luminosity GRBs that may be driven by jets~\citep[e.g.,][]{Margutti:2014gha}.   

We consider an SN shock that propagates in the CSM. For simplicity\footnote{The CSM profile may not be a simple power law, and the shell-like structure has often been observed~\citep[e.g.,][]{Margutti:2016wyh}. In general, we may not apply self-similar solutions to describe the shock dynamics.}, we assume an CSM density profile of
\begin{equation}
\varrho_{\rm cs}(R)={\mathcal D}R^{-w}\equiv DR_{0}^{-2}{\left(\frac{R}{R_{0}}\right)}^{-w},
\label{csm1}
\end{equation}
where $w=2$ corresponds to the wind density profile. The target nucleon density is given by
\begin{eqnarray}
n_{N}=\frac{\mathcal D}{m_H}R^{-w}&\equiv&\frac{D}{m_HR_0^2}{\left(\frac{R}{R_0}\right)}^{-w}~\nonumber\\
&\simeq&3.0\times{10}^{10}~{\rm cm}^{-3}~D_*R_{0,15}^{w-2}R_{15}^{-w},
\end{eqnarray}
where $D_*=D/(5\times{10}^{16}~{\rm g}~{\rm cm}^{-1})$ ($D=\mathcal D$ for $w=2$) and $R_0$ is an arbitrary radius characterizing the CSM radius. 
Note that $D_*=1$ corresponds to $D=\dot{M}_w/(4\pi V_w)$ with a wind mass-loss rate of $\dot{M}_w=0.1~{\rm M}_\odot~{\rm yr}^{-1}$ and a wind velocity of $V_w=100~{\rm km}~{\rm s}^{-1}$. 

As noted, a faster component of the ejecta is decelerated earlier, and the kinetic luminosity that can be used for dissipation at the forward shock is, 
\begin{eqnarray}
L_s=\frac{\Delta\Omega}{2}\varrho_{\rm cs}R_s^2 V_s^3&\simeq&3.9\times{10}^{43}~{\rm erg}~{\rm s}^{-1}~D_*R_{0,15}^{w-2}\nonumber\\
&\times&R_{s,15}^{2-w}{(V_s/5000~{\rm km}~{\rm s}^{-1})}^3,\,\,\,\,\,\,\,\,\,\,
\end{eqnarray}
where $R_s$ is the shock radius, $V_s$ is the forward shock velocity, and $\Delta\Omega$ is the effective solid angle of the CSM interaction.
In this subsection, $\Delta\Omega=4\pi$ is used assuming that the CSM is spherical.   
The bolometric, thermal radiation luminosity (used for thermal SN emission in the optical and/or X-ray bands) is expressed as $L_{\rm rad}=\epsilon_{\rm rad} L_s$, where $\epsilon_{\rm rad}$ is the energy fraction carried by the radiation. 

The neutrino and gamma-ray emission is governed by the effective optical depth of inelastic $pp$ interactions, $f_{pp}$, which is estimated to be
\begin{eqnarray}
f_{pp}&\approx&\kappa_{pp} \sigma_{pp}c n_N (R_s/V_s)\nonumber\\
&\simeq&27~D_*R_{0,15}^{w-2}R_{s,15}^{1-w}{(V_s/5000~{\rm km}~{\rm s}^{-1})}^{-1},
\end{eqnarray}
where $\sigma_{pp}\approx3\times{10}^{-26}~{\rm cm}^2$ is the inelastic $pp$ cross section and $\kappa_{pp}\approx0.5$ is the proton inelasticity, respectively.  Here we have used $f_{pp}\approx\kappa_{pp} \sigma_{pp}c (r_{\rm sc} n_N)(\Delta_s/V_s)\approx\kappa_{pp} \sigma_{pp}cn_N(R_s/V_s)$, where $r_{\rm sc}$ is the shock compression ratio and $\Delta_s\approx R_s/r_{\rm sc}$ is the size of the shocked region. Because the CR energy density in the post-foward-shock region is estimated to be $\epsilon_p\varrho_{\rm cs}V_s^2/2=\epsilon_pL_s/(\Delta\Omega R_s^2 V_s)$, for example, the gamma-ray luminosity is given by
\begin{eqnarray}
L_\gamma &\approx&\frac{1}{3}{\rm min}\left[\frac{M_{\rm cs}(<R_s)}{m_H}\kappa_{pp}\sigma_{pp}c \frac{\epsilon_pL_s}{\Delta\Omega R_s^2 V_s},\epsilon_pL_s\right]\nonumber\\
&\approx& \frac{1}{3}{\rm min}[f_{pp},1]\epsilon_pL_s,
\label{gammapower}
\end{eqnarray}
where $\epsilon_p$ is the energy fraction of accelerated CR protons and $\sim0.1$ is typically expected in the DSA theory for quasi-parallel shocks~\citep{Caprioli:2013dca}. Such a value is also motivated by the hypothesis that interacting SNe are responsible for the observed CR flux around or beyond the knee energy at $\sim3$~PeV~\citep{Sveshnikova:2003sa,Murase:2013kda}. 
The factor $1/3$ comes from $\pi^+:\pi^-:\pi^0\approx1:1:1$ in high-energy inelastic $pp$ interactions.
Note that $M_{\rm cs}(<R_s)=\Delta\Omega DR_s$ in the wind case.  

The principal parameters are CSM nucleon density $n_{N}$, shock radius $R_s$, and shock velocity $V_s$~\citep{Murase:2010cu,Margutti:2013pfa}. The differential gamma-ray luminosity is approximately:
\begin{eqnarray}
E_\gamma L_{E_\gamma}\equiv E_\gamma\frac{d L_\gamma}{dE_\gamma}&\approx&\frac{1}{3}{\rm min}[f_{pp},1]E_pL_{E_p}\nonumber\\
&\simeq&1.3\times{10}^{41}~{\rm erg}~{\rm s}^{-1}~{\rm min}[f_{pp},1]\nonumber\\
&\times&{\left(\frac{E_\gamma}{0.1m_pc^2}\right)}^{2-s_p}\epsilon_{p,-1}{\mathcal R}_{p0,1}^{-1}\nonumber\\
&\times&D_*R_{0,15}^{w-2}R_{s,15}^{2-w}{(V_s/5000~{\rm km}~{\rm s}^{-1})}^3,\,\,\,\,\,\,\,\,
\end{eqnarray}
where $s_p$ is the proton spectral index and ${\mathcal R}_{p0}\sim5-10$ is the normalization factor that is given by
\begin{equation}
E_pL_{E_p}\equiv E_p\frac{d L_p}{dE_p}=\frac{\epsilon_p L_s}{{\mathcal R}_{p0}}{\left(\frac{E_p}{m_pc^2}\right)}^{2-s_p}.
\end{equation} 
Here $E_p$ is the CR proton energy and $E_pL_{E_p}$ is the CR luminosity per logarithmic energy.

In DSA, CRs are accelerated via scatterings with plasma or magnetohydrodynamic waves, and efficient amplification of the magnetic field in both the upstream and downstream regions is naturally expected for SN shocks.
We parameterize the magnetic field by $U_B\equiv\epsilon_B(\varrho_{\rm cs}V_s^2/2)$ (where $U_B$ is the magnetic energy density), which leads to
\begin{eqnarray}
B&=&{(\epsilon_B 4\pi\varrho_{\rm cs}V_s^2)}^{1/2}\nonumber\\
&\simeq&40~{\rm G}~\epsilon_{B,-2}^{1/2}D_*^{1/2}R_{0,15}^{w/2-1}R_{s,15}^{-w/2}{(V_s/5000~{\rm km}~{\rm s}^{-1})},\,\,\,\,\,\,\,\,
\end{eqnarray}
where $\epsilon_B\sim0.001-0.01$ is assumed. Although such a value is motivated by observations of radio SNe~\citep[e.g.,][]{Chevalier:2005aa,Maeda:2012pv} and the hypothesis that these SNe contribute to the observed CR flux beyond the knee energy, it is highly uncertain whether the maximum CR energy can exceed $\sim10-100$~TeV.  
Indeed, the significant amplification of magnetic fields may occur by various physical reasons. The CSM could be highly turbulent and magnetized, because the violent CSM eruptions may also be accompanied by shocks~\citep{Murase:2013kda}. The CSM may also be highly clumpy~\citep{Smith:2008qn}, in which the turbulent dynamo process could amplify both upstream and downstream fields~\citep[e.g.,][]{Xu:2017wsm}.  
The upstream magnetic field amplification with $\epsilon_B\propto\epsilon_p {\mathcal M}_A^{-1}$ (where ${\mathcal M}_A$ is the Alfv\'enic Mach number) could also be realized by CRs themselves via streaming instabilities~\citep{Caprioli:2014tva}. 
The CR spectrum can also be affected by neutral particles~\citep{Murase:2010cu}, and the maximum energy can be limited by the ionized region as an escape boundary (whose size is roughly $\sim0.5\times{10}^{15}~{\rm cm}~{(1+\tau_T)}^{-1/3}L_{X,43}^{1/3}\nu_{18}^{-1/3}n_{N,10}^{-2/3}{({\mathcal T}_{\rm cs,5}^{u})}^{1/4}$ for the intrinsic X-ray luminosity $L_X$ and the pre-shock CSM temperature ${\mathcal T}_{\rm cs}^{u}$). 
We simply determine the maximum CR energy by comparing the acceleration time with the dynamical time and energy-loss time. 
This should be sufficient because our results on GeV-TeV emission are largely insensitive to the maximum energy.  

Secondary electrons and positrons lose their energies via synchrotron cooling, and their characteristic frequency is:
\begin{eqnarray}
\nu_h\sim\frac{3}{4\pi}{\left(\frac{m_\pi}{4m_e}\right)}^2\frac{eB}{m_ec}&\simeq&780~{\rm GHz}~\epsilon_{B,-2}^{1/2}D_*^{1/2}R_{0,15}^{w/2-1}\nonumber\\
&\times&R_{s,15}^{-w/2}{(V_s/5000~{\rm km}~{\rm s}^{-1})}.
\end{eqnarray}
Assuming the fast cooling (i.e., electrons at the injection frequency cool within the dynamical time), the resulting synchrotron luminosity from CR-induced electrons and positrons (at $\nu<\nu_h$) is:
\begin{eqnarray}
\nu L_\nu^h&\approx&\frac{1}{2(1+Y)}\frac{1}{6}{\rm min}[f_{pp},1]E_pL_{E_p}\nonumber\\
&\simeq&3.3\times{10}^{40}~{\rm erg}~{\rm s}^{-1}~{\rm min}[f_{pp},1]{(1+Y)}^{-1}\nonumber\\
&\times&{\left(\frac{\nu}{\nu_h}\right)}^{1/2}\epsilon_{p,-1}{\mathcal R}_{p0,1}^{-1}\nonumber\\
&\times&D_*R_{0,15}^{w-2}R_{s,15}^{2-w}{(V_s/5000~{\rm km}~{\rm s}^{-1})}^3f_{\rm esc},
\end{eqnarray}
where $Y$ is the inverse-Compton Y parameter and $f_{\rm esc}$ is the escape fraction of radio waves. The escape fraction is phenomenologically introduced to represent effects of various low-energy photon absorption processes~\citep{Murase:2013kda}. 

Follow-up observations at high-frequency radio bands are important~\citep{Murase:2013kda,Petropoulou:2016zar}, and the hadronic scenario predicts
\begin{eqnarray}
\frac{E_\gamma F_{E_\gamma}}{\nu F_{\nu}^h}&\approx&4{\left(\frac{E_\gamma}{0.1m_pc^2}\right)}^{2-s_p}{\left(\frac{\nu}{\nu_h}\right)}^{1/2}\nonumber\\
&\sim&0.9~{\left(\frac{E_\gamma}{1~{\rm GeV}}\right)}^{2-s_p}
{\left(\frac{\nu}{100~{\rm GHz}}\right)}^{1/2}\epsilon_{B,-2}^{-1/4}D_*^{-1/4}\nonumber\\
&\times&R_{0,15}^{1/2-w/4}R_{s,15}^{w/4}{(V_s/5000~{\rm km}~{\rm s}^{-1})}^{-1/2}f_{\rm esc}^{-1},
\end{eqnarray}
where $E_\gamma F_{E_\gamma}$ and $\nu F_{\nu}^h$ are gamma-ray and radio energy fluxes observed at Earth, respectively.
Here the numerical value in the last expression is evaluated for $s_p=2.2$ and $Y\ll1$ is assumed. As long as $s_p\sim2$, the ratio of the gamma-ray to submm energy fluxes is expected to be a weak function of time, and is predicted to be the order of unity for Type IIn SNe with $D_*\sim0.1-10$.  
The CR index at sufficiently high energies could be modified in radiative shocks, because a higher compression ratio (i.e., $r_s\gg4$) could make the spectrum harder~\citep[e.g.,][]{Yamazaki:2006uf}. 

As commonly expected in ordinary radio SNe, primary electrons will also be accelerated. The observed CR proton and electron fluxes on Earth and the measurements of Galactic SNRs suggest the ratio of electron to proton fluxes at the same energy is $\sim{10}^{-3}-{10}^{-2}$, implying that the energy fraction carried by DSA-accelerated CR electrons is $\epsilon_e\sim{10}^{-4}-{10}^{-3}$ for a flat energy spectrum~\citep[e.g.,][]{Katz:2007ds}.  Electrons can be injected into the DSA once their Lorentz factor exceeds the shock transition layer, and we assume that the corresponding characteristic Lorentz factor is given by $\gamma_{l2}\sim(m_p/m_e)(V_s/c)$, above which the spectral index is $s_e=s_p$~\citep[e.g.,][]{Park:2014lqa}. 
For $\gamma_{l2}\gg1$, the characteristic synchrotron frequency for the DSA is estimated to be
\begin{eqnarray}
\nu_{l2}=\frac{3}{4\pi}\gamma_{l2}^2\frac{eB}{m_ec}&\sim&160~{\rm GHz}~\epsilon_{B,-2}^{1/2}D_*^{1/2}R_{0,15}^{w/2-1}\nonumber\\
&\times&R_{s,15}^{-w/2}{(V_s/5000~{\rm km}~{\rm s}^{-1})}^3.
\end{eqnarray}
Note that the conventional DSA may not be applied for electrons with lower Lorentz factors. It has been suggested that observations of SNe IIb and SNe IIn will enable us to probe electron acceleration in the shock transition layer~\citep{Maeda:2012pv,Maeda:2012we,Murase:2013kda}. For a low-energy component of non-thermal electrons, the injection Lorentz factor is given by $\gamma_{l1}={[1+{(p_{l1}/m_ec)}^2]}^{1/2}$, where $p_{l1}$ is the corresponding injection momentum. If $p_{l1}\gg m_ec$, we have the conventional formula, $\gamma_{l1}=[(g_e\tilde{\epsilon}_em_p)/(2f_em_e)]{(V_s/c)}^2$, where $\tilde{\epsilon}_e$ is the energy fraction of electrons with $\gamma_e\geq \gamma_{l1}$, $f_e$ is the number fraction, and $g_e=(q_{e}-2)/(q_{e}-1)$ for a low-energy spectral index, $q_{e}>2$. We typically expect $p_{l1}\lesssim m_ec$, where one may approximate a low-energy electron spectrum as $dn_{{\rm CR}e}/d\gamma_e\propto{\gamma_e}^{-q_e}$ with $\gamma_{l1}=1$.   
The steady-state index of electrons can be $\sim3$ due to radiative energy losses, in which $q_e\sim s_e$ is possible~\citep{Chevalier:2005aa}. 
Or, perhaps, $q_e>s_e$ (allowing $\tilde{\epsilon}_e\sim0.01$) could be realized by other processes such as the shock-drift acceleration mechanism at quasi-perpendicular shocks~\citep[e.g.,][]{Matsumoto:2017yag}.   

The high-frequency synchrotron luminosity (at $\nu>\nu_{l2}$) in the fast cooling case is 
\begin{eqnarray}
\nu L_\nu^{l2}&\approx&\frac{1}{2(1+Y)}E_eL_{E_e}\nonumber\\
&\simeq&2.0\times{10}^{38}~{\rm erg}~{\rm s}^{-1}~{(1+Y)}^{-1}{\left(\frac{\nu}{\nu_{l2}}\right)}^{\frac{2-s_e}{2}}\epsilon_{e,-4}{\mathcal R}_{e0,1}^{-1}\nonumber\\
&\times&D_*R_{0,15}^{w-2}R_{s,15}^{2-w}{(V_s/5000~{\rm km}~{\rm s}^{-1})}^3f_{\rm esc},
\end{eqnarray}
where $\epsilon_e(<\tilde{\epsilon}_e)$ is the energy fraction of DSA-accelerated electrons.
Thus, the leptonic origin of radio emission predicts:  
\begin{eqnarray}
\frac{E_\gamma F_{E_\gamma}}{\nu F_{\nu}^{l2}}&\approx&\frac{2\epsilon_p}{3\epsilon_e}\frac{{\mathcal R}_{e0}}{{\mathcal R}_{p0}}{\left(\frac{E_\gamma}{0.1m_pc^2}\right)}^{2-s_p}{\left(\frac{\nu}{\nu_{l2}}\right)}^{\frac{s_e-2}{2}}{\rm min}[f_{pp},1]\nonumber\\
&\sim&400~{\left(\frac{E_\gamma}{1~{\rm GeV}}\right)}^{2-s_p}
{\left(\frac{\nu}{100~{\rm GHz}}\right)}^{\frac{s_e-2}{2}}({10}^{-3}\epsilon_p/\epsilon_e)\nonumber\\
&\times&{({\mathcal R}_{e0}/{\mathcal R}_{p0})}\epsilon_{B,-2}^{\frac{2-s_e}{4}}D_*^{\frac{2-s_e}{4}}R_{0,15}^{\frac{(w-2)(2-s_e)}{4}}R_{s,15}^{\frac{w(s_e-2)}{4}}
\nonumber\\
&\times&{(V_s/5000~{\rm km}~{\rm s}^{-1})}^{\frac{3(2-s_e)}{2}}{\rm min}[f_{pp},1]f_{\rm esc}^{-1},
\end{eqnarray}
where $Y\ll1$ is assumed. The ratio depends on $R_s$ and $V_s$ differently from that in the hadronic scenario, so the time-dependence of the gamma-ray to radio fluxes is relevant to discriminate between the hadronic and leptonic interpretations. 
So far, we have ignored various processes that can suppress the radio emission. More generally, low-frequency emission is affected by various effects such as synchrotron self-absorption, free-free absorption, and Coulomb cooling, which will be included in our numerical calculations. 

In Type IIn SNe, $\epsilon_p\gg\epsilon_e$ allows us to expect that both gamma-ray emission and high-frequency radio emission are dominated by the hadronic component. In the high-frequency limit (i.e., $\nu>\nu_h,\nu_{l2}$), in which either hadronic or leptonic scenario predicts the same spectrum, $\nu F_\nu\propto \nu^{2-\beta}$ (where $\beta$ is the photon index), the condition for the secondary emission to overwhelm the primary component is written:
\begin{equation}
{\rm min}[1,f_{pp}]\frac{\epsilon_{p,-1}}{\epsilon_{e,-4}}{\left(\frac{\nu_{l2}}{\nu_{h}}\right)}^{2-\beta}\gtrsim6\times{10}^{-3},
\end{equation}
where ${\mathcal R}_{p0}={\mathcal R}_{e0}$ and $s=s_p=s_e$ are assumed, and $\beta=1+s/2$ is the high-energy synchrotron photon index predicted by theory\footnote{We note that \cite{Murase:2013kda} used $\epsilon_e={10}^{-4}$ in Eqs.~(58) and (59) and $q$ should be the photon index there.}.
Here $s$ is the spectral index of CR protons and electrons that are accelerated by DSA.

In astrophysical environments, high-energy neutrinos should be produced by hadronic processes like the $pp$ reaction, so they serve as a unique, powerful probe of CR ion acceleration. On the other hand, electromagnetic emissions originate from both hadronic and leptonic processes. In the hadronic scenario, gamma rays are mainly produced by neutral pions and cascades from secondary electrons and positrons, while radio emission is attributed to synchrotron radiation from the secondaries. In the leptonic scenario, gamma rays mainly originate from inverse-Compton radiation by primary electrons, while radiation emission is ascribed to synchrotron radiation from the same primary electrons. In realistic situations, we should expect both components, but the hadronic component is likely to be dominant when DSA occurs in dense environments~\citep{Murase:2013kda}. 
 
\subsection{Phenomenological Model}
In cases of well-observed Type IIn SNe such as SN 2010jl, parameters on SN dynamics can be determined by the observational data of (mostly) thermal emission from the SNe. We here focus on the forward shock emission~\citep[see][for a discussion on the reverse shock emissions]{Chevalier:2016hzo,Murase:2010cu}, and consider a single radiation zone with the size $\Delta R_s\approx R_s$. In a time-dependent model, we need to know three quantities, $\varrho_{\rm cs}$, $R_s$, and $V_s$ as a function of time $t$. If all three quantities are described by power laws, which are assumed throughout this work for simplicity, one needs five parameters, $D$, $w$, $R_{0}$, $t_{0}$, and the temporal index $a$ to describe the SN dynamics. Here, just for convenience, we take $R_{0}=R_{\rm bo}$ and $t_0=t_{\rm bo}$, where $R_{\rm bo}$ is the photon breakout radius and $t_{\rm bo}$ is the breakout time. Our formalism enables us to predict non-thermal fluxes from the observational data only with a few free parameters such as $\epsilon_p$ and $s_p$. 
Note that the calculation framework used in \cite{Murase:2017pfe} and this work are applicable to any type of SNe that cause interactions with the CSM.  For example, not only Type IIn and Type Ibn can be the sources of high-energy neutrinos and gamma rays. High-energy emissions from CSM interactions for Type II-P, II-L and IIb SNe were first studied by \cite{Murase:2017pfe}.  SNe IIn (whose optical emission is powered by the CSM interactions) are typically expected to have the most powerful non-thermal emissions among these SN classes, and extragalactic objects are detectable (see Section~5 for details). The CSM mass indicated for SNe IIn can be as large as $M_{\rm cs}\sim(1-10)~M_{\rm cs}$.

The CSM density given by Equation (\ref{csm1}) can be re-written:
\begin{equation}
\varrho_{\rm cs}(R)=DR_{\rm bo}^{-2}{\left(\frac{R}{R_{\rm bo}}\right)}^{-w},
\end{equation}
where $R_0=R_{\rm bo}$ is used. The CSM nucleon density is given by $n_N=\varrho_{\rm cs}/m_H$. 

Assuming a power-law evolution, the shock radius and velocity are parameterized as
\begin{equation}
R_s=R_{\rm bo}{\left(\frac{t}{t_{\rm bo}}\right)}^{a}
\end{equation} 
and
\begin{equation}
V_s=V_{\rm bo}{\left(\frac{t}{t_{\rm bo}}\right)}^{a-1},
\end{equation}
where $V_{\rm bo}=aR_{\rm bo}/t_{\rm bo}$. 
Note that $R_s$ and $V_s$ are determined by the observational data. 
If the self-similar solution for a spherical CSM is adopted, the index $a$ is explicitly given by $a=(\delta-3)/(\delta-w)$, where $\delta\sim7-12$ is the index of the outer ejecta profile, $\varrho_{\rm ej}\propto t^{-3}{(r/t)}^{-\delta}$~\citep[e.g.,][]{Matzner:1998mg}. 

The shock power is calculated as
\begin{equation}
L_{s}=\left(\frac{\Delta\Omega}{2}\right)D V_{\rm bo}^3{\left(\frac{t}{t_{\rm bo}}\right)}^{5a-3-aw}\propto t^{-\alpha}, 
\label{shockpower}
\end{equation}
where $\alpha=3-a(5-w)$. If we assume the self-similar solution, which is valid until the deceleration of the inner ejecta starts, we obtain $\alpha=[(\delta-3)(w-2)+3(3-w)]/(\delta-w)$~\citep[e.g.,][]{Moriya:2013hka,Ofek:2013afa,Chevalier:2016hzo,Murase:2017pfe}. Note that $\Delta\Omega=4\pi$ in the spherical geometry of the CSM, and it has been argued that the CSM could be aspherical or even clumpy~\citep[e.g.,][]{Smith:2008qn,Margutti:2013pfa,Smith:2014txa,Katsuda:2016zbw}. Only a fraction ($\epsilon_{\rm rad}<1$) of the shock power is used for the radiation luminosity, $L_{\rm rad}=\epsilon_{\rm rad}L_s$.

We assume that the DSA works as in SNRs, and consider a power-law form,
\begin{equation}
\frac{dn_{{\rm CR}p}}{dp}\propto{p}^{-s_p}e^{-p/p_{\rm max}},
\end{equation}
where $n_{{\rm CR}p}$ is the CR proton number density and $dn_{{\rm CR}p}/dp$ is the differential momentum distribution. 
The maximum momentum, $p_{\rm max}$, is determined by CR escape and cooling processes such as $pp$ interactions, Bethe-Heitler pair production, and adiabatic losses\footnote{At sufficiently high energies, the wave damping due to neutral-ion collisions in the upstream region can be relevant as in partially-ionized shocks of SNRs~\citep{Murase:2010cu}, but GeV-TeV gamma-ray and neutrino emission would not be much affected.}. 
We also introduce an escape boundary of $0.6R_s$ to estimate the maximum energy. The escape boundary is assumed to be $\sim(0.1-1)R_s$ for SNRs to explain the observed CR flux~\citep{Ohira:2009rd}, and our results are insensitive to its exact value. 
The DSA thoery for a quasi-parallel shock predicts $s_p=2.0$, whereas a steeper spectrum of $s_p\sim2.2-2.4$ is inferred by GeV-TeV gamma-ray observations of the young SNR Cas A~\citep[e.g.,][and references therein]{Ahnen:2017uny}. 

The normalization of the CR spectrum is given by
\begin{equation}
U_p=\int^{p_{\rm max}} dp \,\,\, E_p \frac{dn_{{\rm CR}p}}{dp},
\end{equation}
where the CR proton energy density $U_p\approx \epsilon_pL_s/(\Delta\Omega R_s^2V_s)$ is used. Throughout this work, we choose $\epsilon_p=0.05$ as a fiducial parameter, based on the observation of Cas A~\citep{Ahnen:2017uny}.
We only consider fresh CRs because the CRs lose their energies via adiabatic losses during the dynamical time (as well as via other energy-loss processes). A more quantitative treatment is possible by solving kinetic equations of protons, which is important only if the luminosity declines more rapidly than $L_s \propto t^{-1}$. 
Our assumption to consider the fresh CRs is justified for $w<3$, and indeed seen in intensively studied SNe IIn~\citep{Moriya:2013hka}. 
Such an approximation is often used in the context of GRBs~\citep{Meszaros:2006rc,Murase:2008sp}.  
Importantly, when the hadronic cooling of CR protons is dominant and its effective time scale, $t_{pp}\approx{(\kappa_{pp}\sigma_{pp}n_Nc)}^{-1}$, is shorter than the dynamical time, $t_{\rm dyn}\approx R_s/V_s$, the luminosity of secondary particles such as neutrinos and gamma rays is limited by $L_{p}$ (see Equation~\ref{gammapower}). The system is ``calorimetric" and energy conservation should not be violated. 

For a given set of parameters describing hydrodynamics evolution of the parameters, we calculate energy densities of neutrinos, gamma rays, and electrons/positrons, by exploiting the following formulas:
\begin{eqnarray}
\dot{n}_{E_\nu}^{\rm inj}=\frac{d\sigma_{pp}\xi_\nu}{dE_\nu} \frac{cM_{\rm cs}}{m_H\mathcal V} \int^{p_{\rm max}} dp \,\,\, E_p \frac{dn_{{\rm CR}p}}{dp}\nonumber\\
\dot{n}_{E_\gamma}^{\rm inj}=\frac{d\sigma_{pp}\xi_\gamma}{dE_\gamma} \frac{cM_{\rm cs}}{m_H\mathcal V}\int^{p_{\rm max}} dp \,\,\, E_p \frac{dn_{{\rm CR}p}}{dp}\nonumber\\
\dot{n}_{E_e}^{\rm inj}=\frac{d\sigma_{pp}\xi_e}{dE_e} \frac{cM_{\rm cs}}{m_H\mathcal V}\int^{p_{\rm max}} dp \,\,\, E_p \frac{dn_{{\rm CR}p}}{dp},
\end{eqnarray}
where $\xi_\nu$, $\xi_\nu$, and $\xi_e$ represent multiplicities of the secondaries. 
We use the differential cross sections of $pp$ interactions given by \cite{Kelner:2006tc}, and the total $pp$ cross section is adjusted to the post--Large-Hadron-Collider formula by \cite{Kafexhiu:2014cua}. Here the volume of the radiation zone is approximated as $\mathcal V\approx \Delta\Omega R_s^3/3$.  

In the previous work, \cite{Murase:2010cu}, the secondary emissions were calculated ignoring effects of electromagnetic cascades. In this work, using the code that was previously developed in \cite{Murase:2017pfe}, we solve the following kinetic equations: 
\begin{eqnarray}\label{eq:cascade}
\frac{\pd n_{E_e}^e}{\pd t} &=& \frac{\pd n_{E_e}^{(\gamma\gamma)}}{\pd t} 
- \frac{\pd}{\pd E_e} [(P_{\rm IC}+P_{\rm syn}+P_{\rm ff}+P_{\rm ad}+P_{\rm Cou}) n_{E_e}^e]\nonumber\\ 
&+& \dot{n}_{E_e}^{\rm inj},\nonumber\\
\frac{\pd n_{E_\gamma}^\gamma}{\pd t} &=& -\frac{n_{E_\gamma}^{\gamma}}{t_{\gamma \gamma}} -\frac{n_{E_\gamma}^{\gamma}}{t_{\rm mat}} - \frac{n_{E_\gamma}^{\gamma}}{t_{\rm esc}}
+ \frac{\pd n_{E_\gamma}^{(\rm IC)}}{\pd t} 
+ \frac{\pd n_{E_\gamma}^{(\rm syn)}}{\pd t}\nonumber\\ 
&+& \frac{\pd n_{E_\gamma}^{(\rm ff)}}{\pd t}+\dot{n}_{E_\gamma}^{\rm inj},
\end{eqnarray}
where 
\begin{eqnarray}
t_{\gamma \gamma}^{-1} &=& \int d E_\gamma \,\, n_{E_\gamma}^\gamma  \int \frac{d \cos\theta}{2} \,\, \tilde{c} \sigma_{\gamma \gamma} 
, \nonumber \\
\frac{\pd n_{E_\gamma}^{(\rm IC)}}{\pd t} &=& \int d E_e \,\, n_{E_e}^e \, \int d E_\gamma\,\, n_{E_\gamma}^\gamma \, \int  \frac{d \cos\theta}{2} \,\, \tilde{c}   \frac{d \sigma_{\rm IC}}{d E_\gamma} 
, \nonumber \\
\frac{\pd n_{E_e}^{(\gamma \gamma)}}{\pd t} &=& \frac{1}{2} \int d E_\gamma\,\, n_{E_\gamma}^\gamma \, \int d E'_\gamma\,\, n_{E'_\gamma}^\gamma \, \int \frac{d \cos\theta}{2} \,\, \tilde{c}  \frac{d \sigma_{\gamma \gamma}}{d E_e} 
, \nonumber
\end{eqnarray}
and $\pd n_{E_\gamma}^{(\rm ff)}/\pd t$ represents bremsstrahlung emission from non-thermal electrons and positrons~\citep{1975clel.book.....J}. 
Here $\tilde{c} =(1-\cos\theta)c$ (where $\theta$ is the angle between two particles), $t_{\gamma\gamma}$ is the two-photon annihilation time, $t_{\rm mat}$ is the energy-loss time due to Compton scattering and Bethe-Heitler pair-production~\citep{1992ApJ...400..181C} processes, $t_{\rm esc}\approx R_s/c$ is the photon escape time, $P_{\rm IC}$ is the inverse-Compton energy-loss rate (where the Klein-Nishina effect is fully taken into account), $P_{\rm syn}$ is the synchrotron energy-loss rate~\citep{1979rpa..book.....R}, $P_{\rm ff}$ is the non-thermal bremsstrahlung energy-loss rate~\citep{2002cra..book.....S}, $P_{\rm ad}$ is the adiabatic energy-loss rate, and $P_{\rm Cou}$ is the Coulomb energy-loss rate~\citep{2002cra..book.....S}. The calculation during $t_{\rm dyn}\approx R_s/V_s$ essentially leads to quasi-steady photon spectra at $t$. 

We parameterize the magnetic field in the radiation zone by $\epsilon_B\sim{10}^{-3}-{10}^{-1}$ against the ram pressure\footnote{Alternatively, one can introduce the magnetic field as $U_{B}\equiv\varepsilon_{B}{\mathcal E}_s/{\mathcal V}$, where ${\mathcal E}_s\approx L_s(R_s/V_s)$ is the dissipation energy.}.
Note that predictions for high-energy neutrinos depend only on $\epsilon_p$, $s_p$, and $\epsilon_B$, when all parameters on the SN dynamics are determined by the observational data. 

On the other hand, electromagnetic spectra are largely affected by the SN thermal radiation itself and various attenuation processes in the CSM. First, assuming a gray-body spectrum, we introduce the SN optical emission as
\begin{equation}
L_{\rm sn}=L_{\rm sn0}{\left(\frac{t}{t_{\rm bo}}\right)}^{-b_1}
\end{equation}
for the optical luminosity, and
\begin{equation}
{\mathcal T}_{\rm sn}={\mathcal T}_{\rm sn0}{\left(\frac{t}{t_{\rm bo}}\right)}^{-b_2}
\end{equation}
for the temperature. The thermal energy may be efficiently converted into radiation via bremsstrahlung or Compton scattering. If the thermalization efficiently proceeds, a significant fraction of the radiation luminosity is released as the SN emission in the optical band.  Then we expect $L_{\rm rad}\approx L_{\rm sn}=\epsilon_{\rm sn}L_{s}$ with $b_1=\alpha$. Note that the shock power is estimated from the optical luminosity by $L_s=L_{\rm sn}/\epsilon_{\rm sn}$.     
In the SN 2010jl case we discuss below, we also use a gray body spectrum and adopt $b_2=0$ for simplicity. 
The SN thermal radiation energy density at $R_s$ is:
\begin{equation}
U_{\rm sn}\approx\frac{(1+\tau_T)3L_{\rm sn}}{4\pi R_s^2 c}.
\end{equation}
When the two-photon annihilation optical depth in the radiation zone, $\tau_{\gamma\gamma}$, is larger than unity, this external radiation field unavoidably develops an electromagnetic cascade, which can also suppress the gamma-ray spectrum in the TeV range as we see below. 
In addition, we include the X-ray emission originating from thermal bremsstrahlung. The total radiation luminosity from thermal particles is $L_{\rm rad}\approx L_{\rm sn}+L_{X}$. We use $L_X\approx{\rm min}[L_{\rm ff},0.5L_s]$, where $L_{\rm ff}$ is the luminosity of thermal bremsstrahlung emission~\citep[e.g.,][]{1979rpa..book.....R,Murase:2013kda}. The immediate post-shock temperature is determined by the shock velocity $V_s$, which typically lies in the X-ray range~\citep[e.g.,][]{Chevalier:2016hzo}. Note that the observed X-ray luminosity is generally much lower than $L_X$ for the radiation zone, due to the photoelectric absorption in the pre-shock CSM~\citep[e.g.,][]{Chevalier:2016hzo}. Considering the thermalization, the average X-ray energy density is estimated to be $U_X\approx3L_X/(4\pi R_s^2c)/(1+\tau_T)$. However, the inclusion of X rays does not essentially affect the results on non-thermal spectra. 

Next, we consider absorption processes for photons escaping from the radiation zone. The screen region is essentially the far upstream of the forward shock. 
Gamma rays are further attenuated via the two-photon annihilation process by the SN photons scattered in the pre-shock CSM, as well as via Compton scattering, Bethe-Heitler pair-production, and photoelectric absorption by interactions with the matter. Denoting energy-dependent optical depths to each process by $\tau_{\gamma\gamma}^u$, $\tau_{\rm Comp}^u$, $\tau_{\rm BH}^u$, and $\tau_{\rm pe}^u$, we phenomenologically implement the suppression factor, $f_{\rm sup}=\exp(-\tau_{\gamma\gamma}^u)f_{\rm sup}^{\rm mat}$, where $f_{\rm sup}^{\rm mat}=\exp(-\kappa_{\rm Comp}\tau_{\rm Comp}^u-\kappa_{\rm BH}\tau_{\rm BH}^u-\tau_{\rm pe}^u)$, $\kappa_{\rm Comp}$ and $\kappa_{\rm BH}$ are energy-dependent inelasticities~\citep{Murase:2014bfa}. 
A fraction of the SN photons scattered at $R$ is $\sim\tau_T(R)$ and the radiation field drops with the inverse-square law. Ignoring details of the geometrical effect for simplicity, we approximately use $\tau_{\gamma\gamma}^u\approx\tau_T\tau_{\gamma\gamma}{(1+\tau_T)}^{-1}w^{-1}$. 
The cross section of the Bethe-Heitler process depends on the nucleus charge. Throughout this work, we assume a mass composition of $x_{\rm H}=0.7$, $x_{\rm He}=0.25$, and $x_{\rm C/O}=0.05$, leading to the effective charge, $\tilde{Z}_{\rm eff}\equiv\Sigma_i (2Z_i^2/A_i)x_i=2.25$, and the inverse of the mean molecular weight per electron, $\mu_e^{-1}=0.85$.     
At lower energies, the photoionization and excitation become relevant, for which we adopt the simple approximate opacity, $K_{\rm pe}\simeq2.4\times{10}^{-2}~{\rm cm}^2~{\rm g}^{-1}~{(h\nu/10~{\rm keV})}^{-3}$~\citep{1983ApJ...270..119M}. For Type IIn SNe, non-thermal X-ray emission is usually masked by the thermal bremsstrahlung component, and our work has a more focus on gamma rays and radio waves. 

On the other hand, radio emission is affected by the synchrotron self-absorption, free-free absorption by free electrons, and the so-called Razin-Tsytovich effect. We implement $[1-\exp(-\tau_{\rm sa})]/\tau_{\rm sa}$ for the self-absorption in the radiation zone, where $\tau_{\rm sa}$ is calculated for the energy distribution of electrons and positrons, obtained by our numerical calculations~\citep{1979rpa..book.....R}. 
The free-free absorption optical depth, $\tau_{\rm ff}^u$, is:
\begin{eqnarray}
\tau_{\rm ff}(\nu)&\approx&\int _{R_s}dR\, 8.5\times{10}^{-28}~{\left(\frac{\nu}{{10}^{10}~{\rm Hz}}\right)}^{-2.1}n_e^2\nonumber\\
&\times&{\left(\frac{{\mathcal T}_{\rm cs}^u}{10^4~{\rm K}}\right)}^{-1.35}\left(\frac{1-{\rm e}^{-h\nu/k{\mathcal T}_{\rm cs}^u}}{h\nu/k{\mathcal T}_{\rm cs}^u}\right),
\end{eqnarray}
ignoring the detailed dependence on the metallicity. We expect that the immediate upstream region is ionized by X rays, and we use $f_{\rm sup}^{\rm mat}=\exp(-\tau_{\rm ff}^u)$. Note that ${\mathcal T}_{\rm cs}^{u}$ is a pre-shock CSM temperature. Because the far upstream region is expected to be partially ionized or nearly neutral, the above estimate gives conservative results. We also include the Razin-Tsytovich effect simply by $f_{\rm sup}^{\rm mat}=\exp(-\nu_{\rm RT}/\nu)$, where $\nu_{\rm RT}$ is the Razin-Tsytovich frequency. See discussions in \cite{Murase:2013kda} for details of these physical effects.  

The differential luminosity of non-thermal radiation is calculated by 
\begin{equation}
E_\gamma L_{E_\gamma}= \frac{(E_\gamma^2 n_{E_\gamma}){\mathcal V}f_{\rm sup}}{t_{\rm esc}}.
\end{equation}
For a clumpy or aspherical CSM, the attenuation due to matter (not radiation) can be significantly alleviated, which is especially important for the detectability of radio emission. In such a case, one could replace the suppression factor due to interactions with matter ($f_{\rm sup}^{\rm mat}$) with ${\rm max}[f_{\rm sup}^{\rm mat},f_{\rm esc}^{\rm mat}]$, where $f_{\rm esc}^{\rm mat}$ is the photon escape fraction that depends on details of the CSM geometry. Note that $f_{\rm sup}=1$ for neutrinos, and the flavor mixing is taken into account. 

\begin{table*}
\centering
\caption{Source parameters of SN 2010jl, used in our time-dependent model. See text for details~\citep[see also][]{Ofek:2013afa}. The derived shock power corresponds to $\epsilon_{\rm sn}=0.25$.}
\begin{tabular}{|c|c|c|c|c||c|c||c|c|c|}
\hline $D_*$ & $w$ & $t_{\rm bo}$ [d] & $R_{\rm bo}$ [cm] & $a$ & $L_{\rm sn0}$ [erg/s] & ${\mathcal T}_{\rm sn0}$ [K] & $\epsilon_{B}$ & $\epsilon_p$\\
\hline $6$ & $2$ & $20$ & $1.1\times{10}^{15}$ & $0.875$ & $5.5\times{10}^{43}$ & $9000$ & $1.5\times{10}^{-2}$ & $0.05$\\
\hline
\end{tabular}
\end{table*} 

\begin{figure}[tb]
\includegraphics[width=\linewidth]{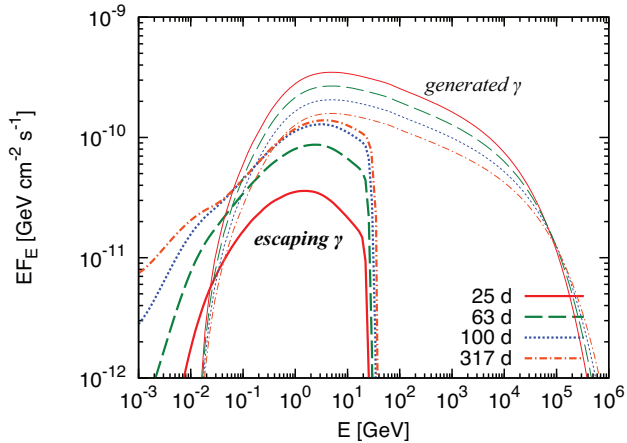}
\caption{Theoretical predictions for hadronic gamma-ray spectra from SN 2010jl at different times, obtained by our time-dependent numerical model. We show cases after (thick curves) and before (thin curves) gamma-ray attenuation in the CSM. Note that electromagnetic cascades in the radiation zone are taken into account. The CR parameters are $\epsilon_p=0.05$ and $s_p=2.2$.
}
\end{figure}
\begin{figure}[tb]
\includegraphics[width=\linewidth]{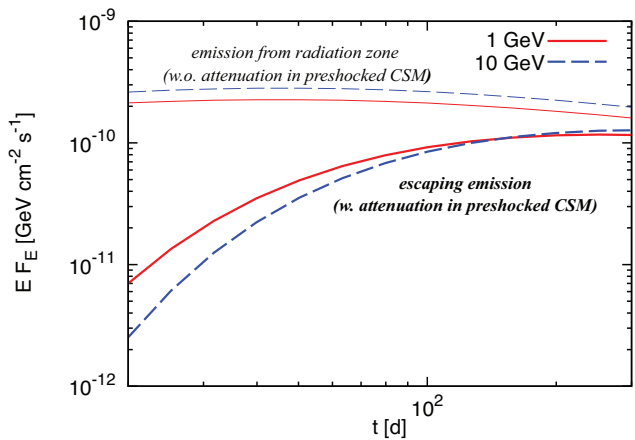}
\caption{Modeled hadronic gamma-ray light curves of SN 2010jl, corresponding to Figures~1 and 2. 
Possible gamma-ray attenuation in the GeV range, indicated by the thick curves, is dominated by the Bethe-Heitler pair-production process in the pre-shock CSM.  
Note that this attenuation is irrelevant for late-time emission that dominates the gamma-ray fluence. 
}
\end{figure}
\begin{figure}
\includegraphics[width=\linewidth]{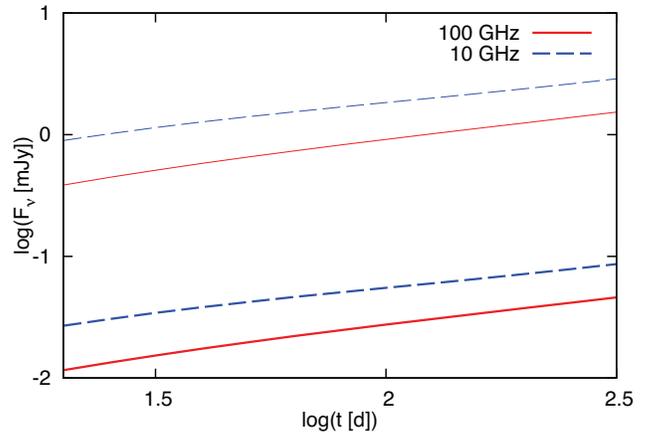}
\caption{Radio light curves corresponding to Figure~3. The attenuation of radio waves during the CSM propagation is too strong. We here show cases for radio emission from the radiation zone (thin curves) and escaping emission for $f_{\rm esc}^{\rm mat}=0.03$ (thick curves). 
}
\end{figure}

\section{Application to SN 2010jl}
\subsection{Theoretical Prediction}
We apply our calculation method to SN 2010jl, adopting the physical parameters estimated by \cite{Ofek:2013afa} (see Table~1). 
SN 2010jl was found in the star-forming galaxy UGC 5189A whose redshift is $z=0.0107$ corresponding to a distance of $d=49$~Mpc~\citep{Ofek:2013afa}. 
The SN coordinates are ${\rm R.A.}={09}^{\rm h}{42}^{\rm m}{53.3}^{\rm s}$ and ${\rm Dec.}={+09}^{\circ}{29}^{'}{42}^{''}$.

Although our results on gamma rays are unaffected by interactions with X rays, we also include thermal X rays as an additional photon field. For simplicity, the X-ray energy density at the source is implemented by using the formula of bremsstrahlung emission with an immediate downstream temperature of $(3/16)\mu m_HV_s^2$, where $\mu$ is the mean molecular weight. (Note that the post-shock temperature can be much lower in the far downstream for a radiation-dominated or radiative shock.) 

The results on modeled gamma-ray spectra and light curves are shown in Figures~2 and 3. We consider hadronic gamma-ray production, and one sees that pionic gamma rays give a dominant contribution. Gamma-ray attenuation by the Bethe-Heitler pair-production process in the CSM is important at early times around the photon breakout time, but at late times the gamma-ray flux is only moderately suppressed below $\sim20-30$~GeV. Note that this effect is less important if the shock velocity is higher~\citep{Murase:2013kda}, and its overall influence on the fluence (to $t\sim300$~d) is only a factor of 2 at most (see Figure~1). This means that GeV gamma rays can be used as a much more direct probe of shock interactions in dense environments such that $\tau_T>1$, than visible light, X rays, and radio waves that are subject to scatterings and attenuation.

In addition, the two-photon annihilation ($\gamma\gamma\rightarrow e^+e^-$) due to interactions with SN photons is crucial above a cutoff energy at $\sim20-30$~GeV, which is consistent with $E_\gamma(3k{\mathcal T}_{\rm sn})\approx m_e^2c^4$. Note that we have assumed a constant SN temperature of ${\mathcal T}_{\rm sn}=9000$~K assuming the gray body radiation. 
Electrons and positrons from muon decay as well as those from $\gamma\gamma\rightarrow e^+e^-$ radiatively cool mainly via synchrotron radiation, but a fraction of their energy is used for inverse-Compton emission, which can be seen below $\sim0.1$~GeV energies. The light curves are shown in Figure~3, where one sees that the gamma-ray attenuation in the CSM can be important until $\sim100$~d, which makes the predictions for gamma-ray light curves different from those without the attenuation effects.    

We consider light curves from $t=20$~d to $t=316$~d. The time window of our theoretical calculation and {\it Fermi}-LAT gamma-ray search is motivated by the sharp decline of the observed optical light curve at $t\sim300$~d. Its origin has been under debate~\citep[e.g.,][]{Ofek:2013afa,Maeda:2013hja}, and there may be a break in the bolometric light curve itself~\citep{Fransson:2013qya}.  The gamma-ray and neutrino luminosities are proportional to ${\rm min}[1,f_{pp}]\epsilon_pL_s$. With $f_{pp}\gtrsim1$, they will trace the bolometric luminosity that could abruptly drop after the shock reaches the edge of the CSM (whose radius is characterized by $R_w$).  

We also calculate radio light curves in Figure~4. However, as expected, radio emission is strongly suppressed due to strong absorption in the CSM as well as synchrotron self-absorption in the radiation zone. In particular, the free-free absorption process is crucial especially when the pre-shock CSM is uniform and spherical, and we do not expect radio detection even if we assume a rather high temperature of ${\mathcal T}_{\rm cs}^{u}=3\times10^5$~K. However, a fraction of the radio emission could escape if the CSM is aspherical or even clumpy, which has been suggested observationally for some Type IIn SNe~\citep{Smith:2014txa,Margutti:2013pfa}. Also, a fraction of escaping radio waves could be enhanced if the pre-shock CSM is mostly neutral. To discuss this effect, we here assume that 3\% of the emission can escape because of the possible incomplete CSM coverage, while the attenuation due to radiation fields is fully taken into account. The radio emission from SN 2010jl had not been detected for about a year, which is consistent with the strong absorption. The radio observations of SN 2010jl were carried out with the Karl G. Jansky Very Large Array, and detections were reported $\sim600-700$~d after the explosion~\citep{Chandra:2015jsa}. Although the detailed modeling of the radio emission is beyond the scope of this work, the radio data show a late time flux with $\sim0.1$~mJy, which is roughly consistent with our predictions of the hadronic scenario if $\sim3$\% of the emission can avoid interactions with electrons in the matter. 

\begin{figure*}
\centering
\includegraphics[width=0.32\linewidth]{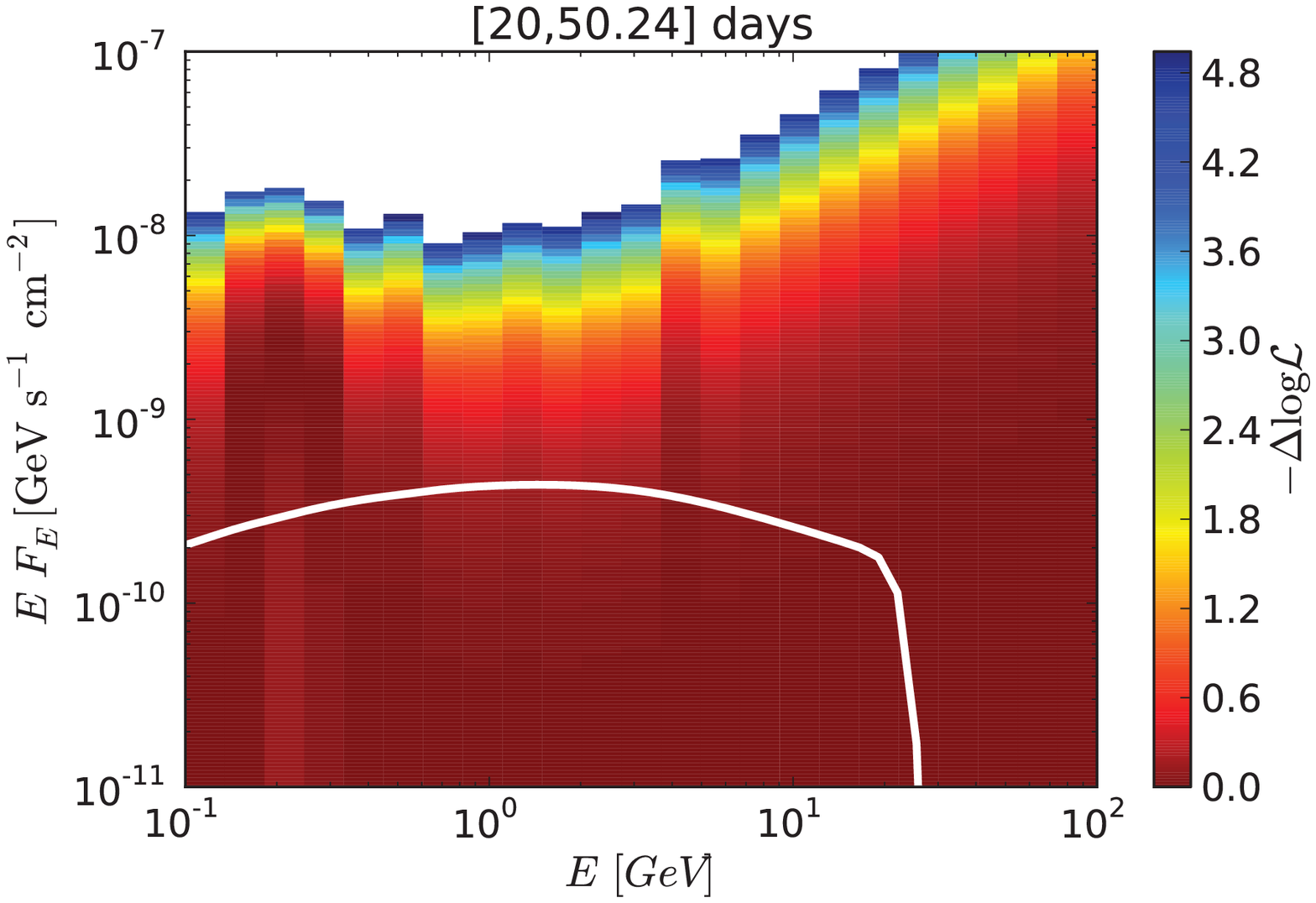}\hfill
\includegraphics[width=0.32\linewidth]{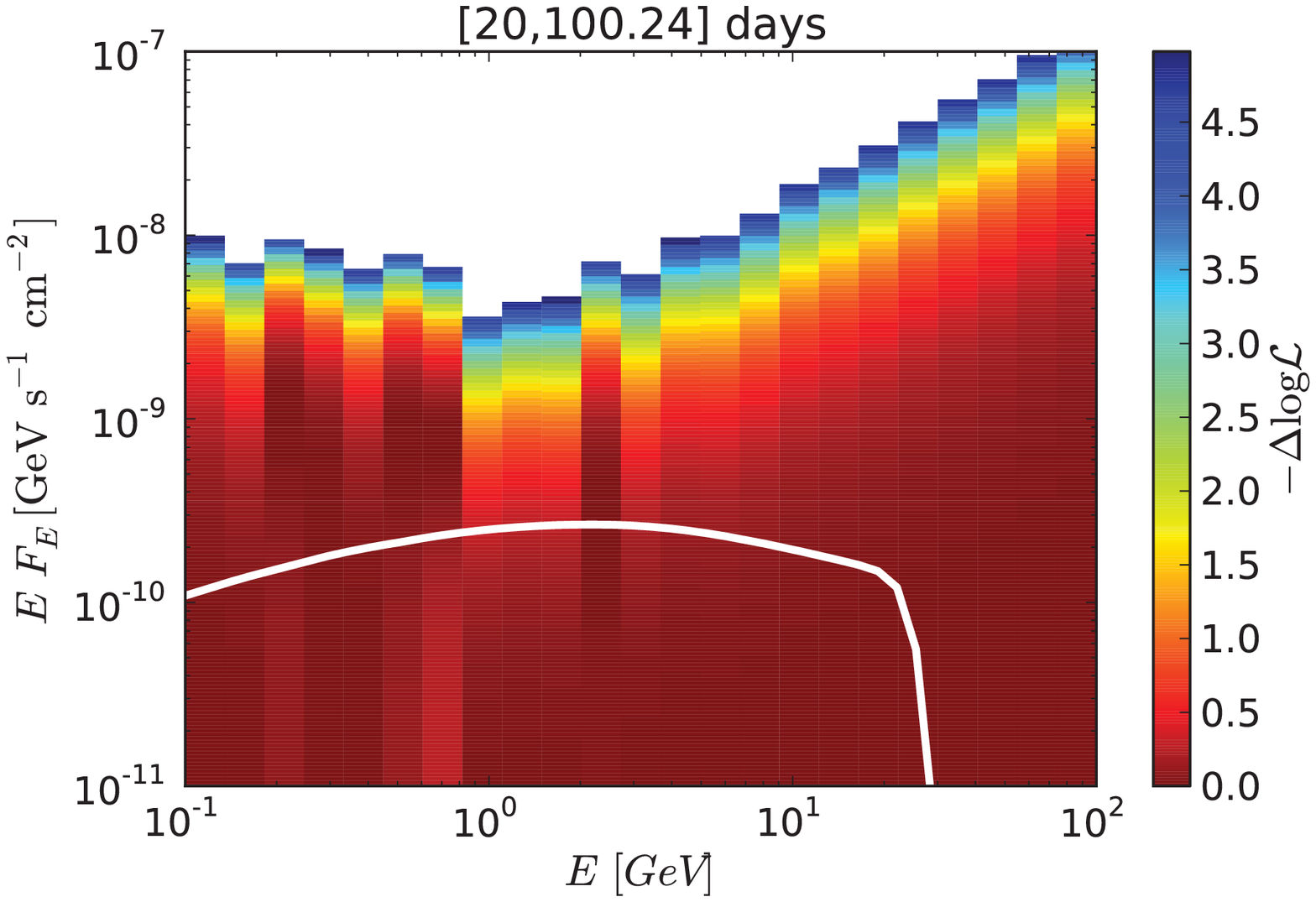}\hfill
\includegraphics[width=0.32\linewidth]{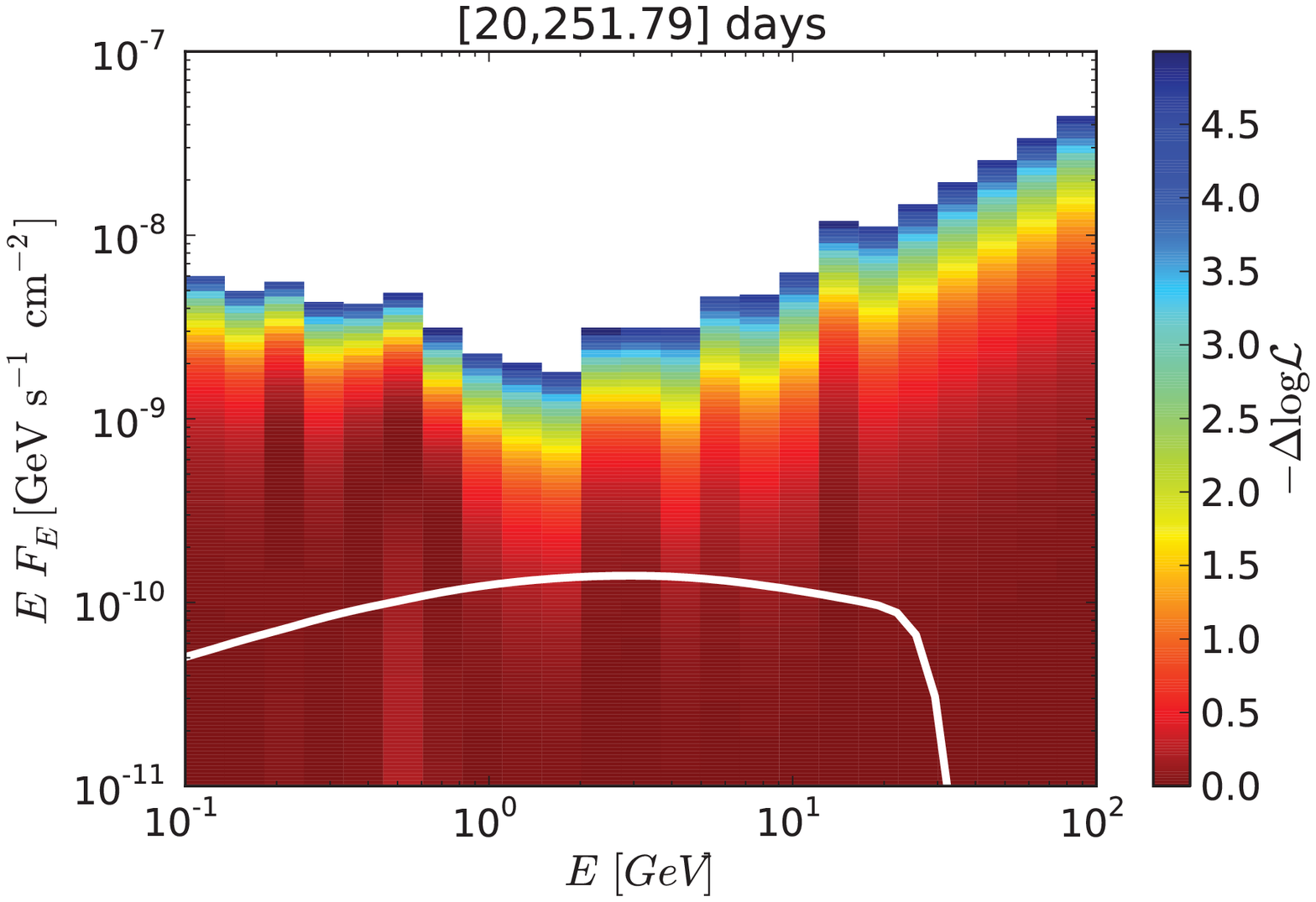}
\caption{{\it Fermi}-LAT constraints on gamma-ray fluxes from SN 2010jl, for different integration times from 20~d to 50.24~d (left), 100.24~d (middle), and 251.79~d (right). Within each energy bin, the color scale denotes the variation of the logarithm of the likelihood with respect to the best-fit value of the SN flux for a given time window. We test a putative source at the SN position and construct the bin-by-bin likelihood function. Then the bin-by-bin likelihood is calculated by scanning the integrated energy flux of the SN within each energy bin. The theoretical expected fluxes for $s_p=2.2$, averaged over time windows, are shown as solid curves (with gamma-ray attenuation in the pre-shock CSM).   
}
\end{figure*}

\begin{figure*}
\centering
\includegraphics[width=0.32\linewidth]{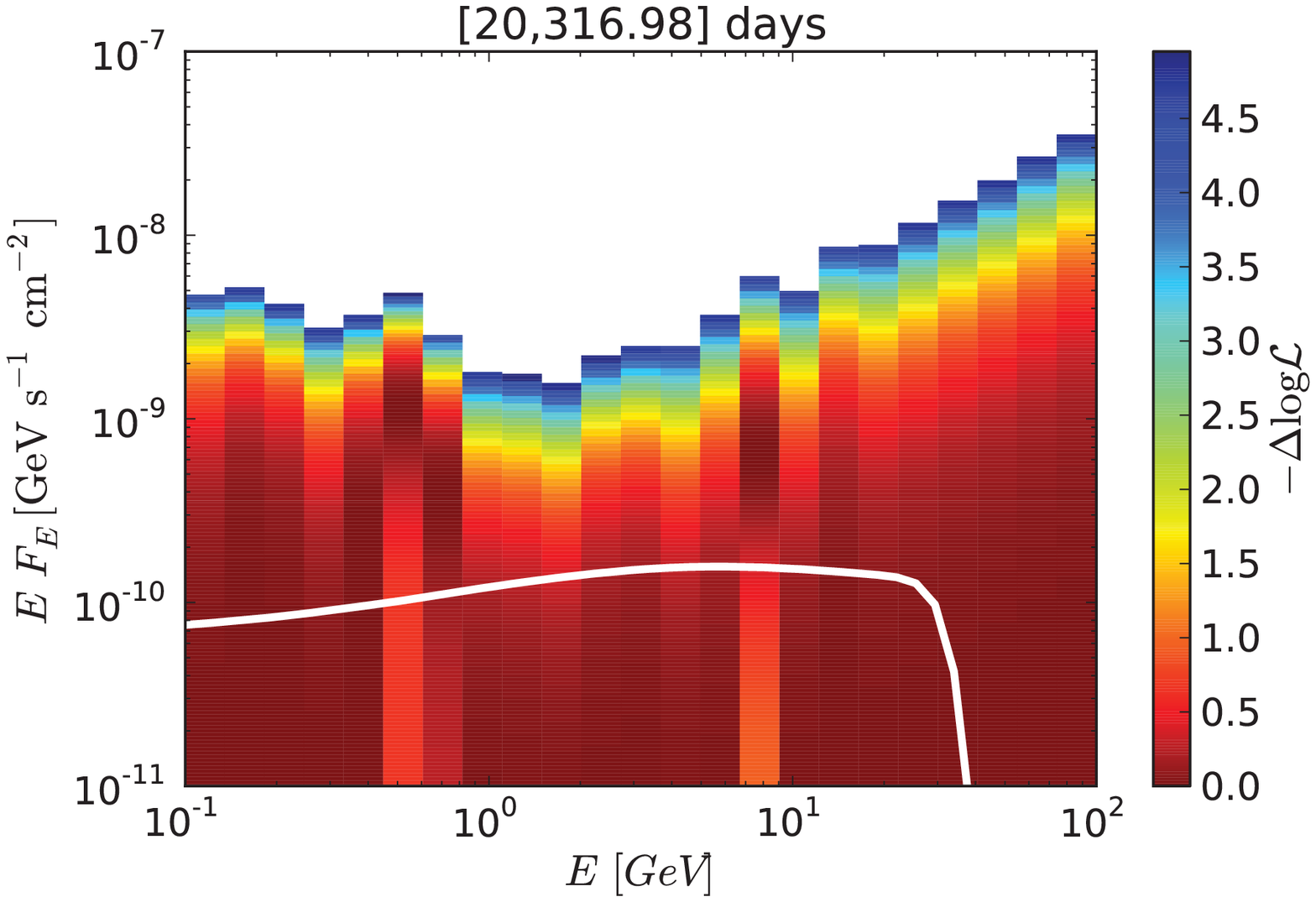}\hfill
\includegraphics[width=0.32\linewidth]{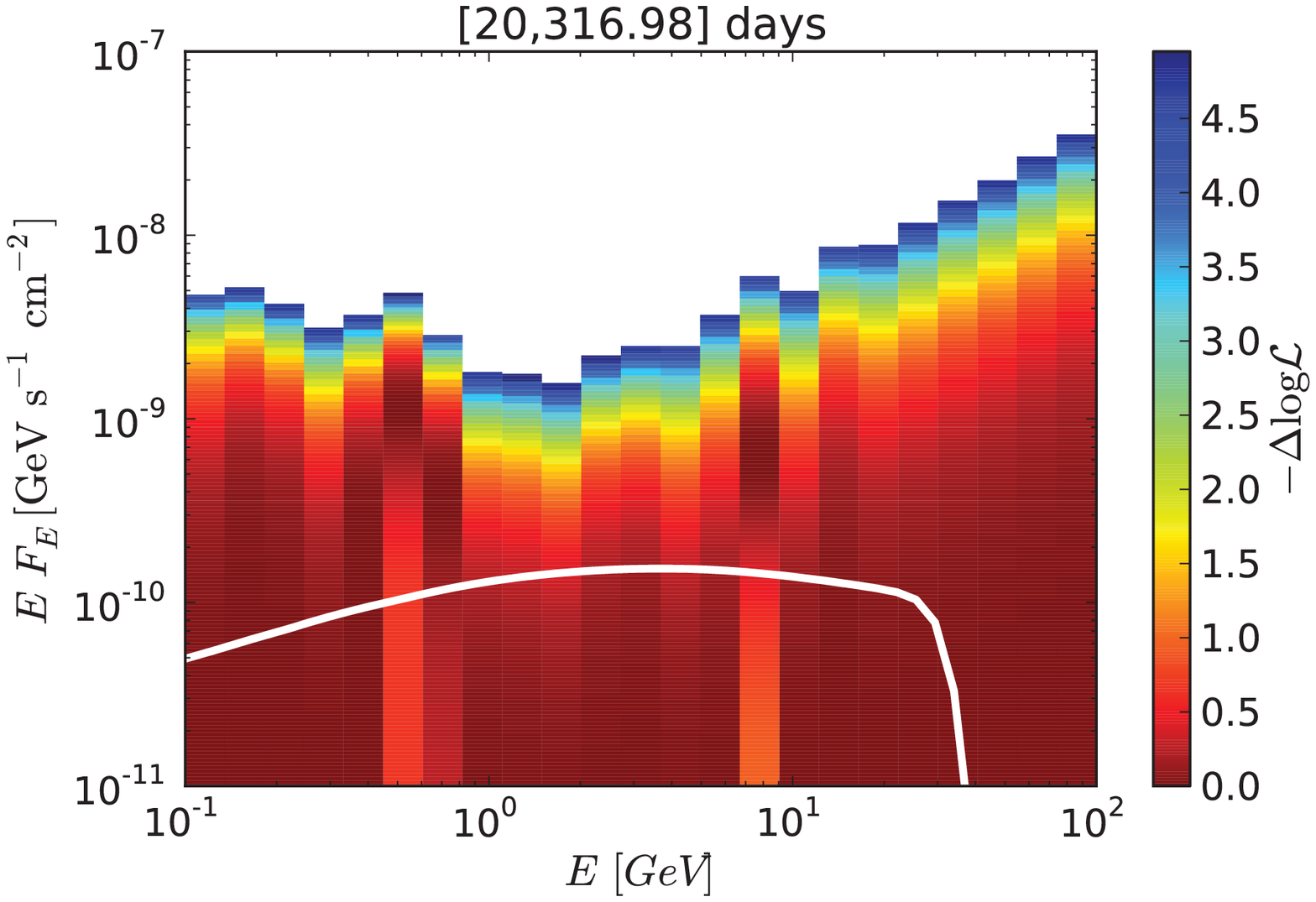}\hfill
\includegraphics[width=0.32\linewidth]{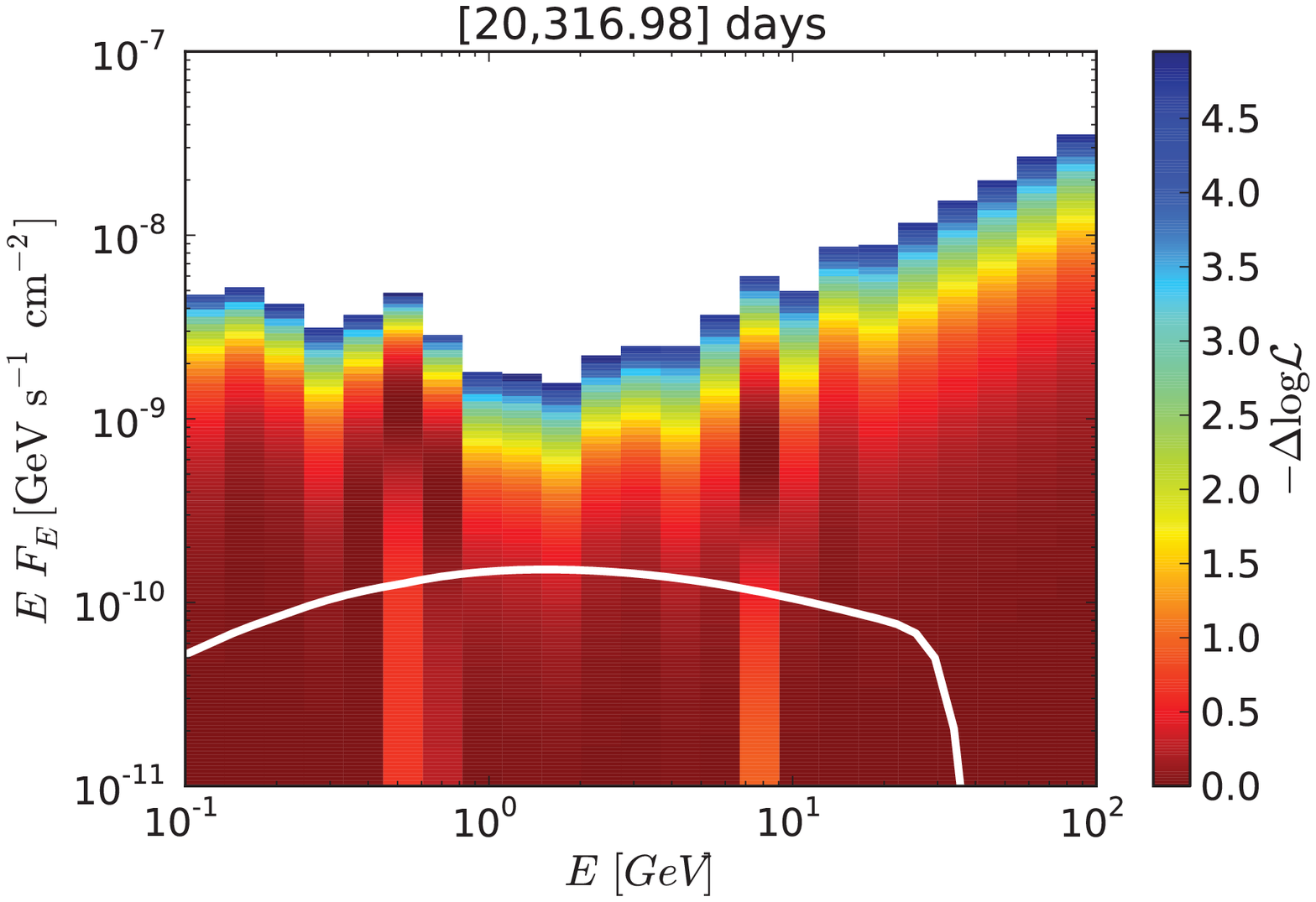}
\caption{{\it Fermi}-LAT constraints on gamma-ray fluxes from SN 2010jl for an integration time from 20~d to 316.98~d. Similar to Figure~5, but the CR spectral index is set to $2.0$ (left), $2.2$ (middle), and $2.4$ (right), respectively.  
}
\end{figure*}

\subsection{Fermi-LAT Data Analysis and Implications}
Dedicated searches for gamma-ray emission from interaction SNe were first performed by \cite{TheFermiLAT:2015kla}. The SN sample used in the past stacking analysis include SN 2010jl. In this work, focusing on SN 2010jl, we re-analyze the {\it Fermi}-LAT data with the Pass 8 SOURCE class\footnote{\url{http://fermi.gsfc.nasa.gov/ssc/data/analysis/
documentation/Pass8_usage.html}}. 
There are two significant improvements compared to the previous work. First, observationally, the Pass 8 data benefit from improved reconstruction and event selection algorithms with respect to the previous data release Pass 7 leading to a significantly improved angular resolution and sensitivity~\citep{Atwood:2009ez} and therefore provide improved gamma-ray constraints. Second, the rich observational data on SN 2010jl allows us to derive detailed physical constraints on the possibility of CR ion acceleration in interacting SNe. In \cite{TheFermiLAT:2015kla}, simple gamma-ray spectra with neither attenuation nor cascades were used, whereas we employ more realistic gamma-ray spectra by including these detailed effects. Also, thanks to the time-dependent multi-wavelength data for the SN, we are able to convert the gamma-ray limits into the bounds on the CR energy fraction, which is the quantity of particular interest.   

We perform a binned analysis (i.e., binned in space and energy) using the standard {\it Fermi}-LAT Science-Tools package version v10r01p01 available from the Fermi Science Support Center\footnote{\url{http://fermi.gsfc.nasa.gov/ssc/data/analysis/}} (FSSC) and the P8R2\_SOURCE\_V6 instrument response functions. Otherwise the analysis is identical to the one described in \cite{TheFermiLAT:2015kla}. 
The results of our updated gamma-ray data analysis at different integration times are shown in Figure~5.  We here plot the logarithm of the likelihood ratio, $-\Delta\ln {\mathcal L}\equiv \ln({\mathcal L}/{\mathcal L}_0)$, as a function of energy. Here ${\mathcal L}_0$ is the likelihood evaluated at the best-fit parameters under a background-only hypothesis, whereas ${\mathcal L}$ is the likelihood evaluated at the best-fit model parameters with a candidate point source at the SN position. 
See \cite{TheFermiLAT:2015kla} for details on the likelihood analysis. 

Both differential and spectrum-integrated upper limits on gamma-ray fluxes averaged over a given time window are shown. At lower energies, the flux sensitivity approximately scales as $\propto t^{-1/2}$, as expected in the background-dominated regime. At high energies, it scales as $\propto t^{-1}$ at early times and then becomes $\propto t^{-1/2}$.  
Note that the spectrum-integrated flux limits that depend on theoretical flux templates are typically several times stronger than the differential flux limits, for a hard spectrum with $s\sim2$.   

Based on these results, we place quantitative limits on the energy fraction of CR protons, $\epsilon_{p}$. This is done by convolving the gamma-ray spectra obtained by the theoretical calculations (Section~3) and the differential upper limits shown in Figures~5 and 6.  
For different integration times from 20~d to 50.24~d, 100.24~d, and 251.79~d, the derived 95\% CL limits for $s_p=2.2$ are $\epsilon_p\leq0.75~(\epsilon_{\rm sn}/0.25)$, $\epsilon_p\leq0.21~(\epsilon_{\rm sn}/0.25)$, and $\epsilon_p\leq0.066~(\epsilon_{\rm sn}/0.25)$, respectively. 
See Figure~1 for the comparison between our theoretical predictions and the gamma-ray limit from the {\it Fermi}-LAT data. 
In Figure~1, the possible effect of gamma-ray attenuation in the pre-shock CSM is also indicated, and the constraints on $\epsilon_p$ can be somewhat improved if the gamma-ray attenuation in the CSM is completely negligible.    
Note that $\epsilon_p=0.05$, which is indicated by the observation of the Galactic SNR Cas A, is still consistent with the non-detection of gamma rays from this SN IIn, and the upper limit can be somewhat weaker for larger values of $\epsilon_{\rm sn}$ (as we are constraining $\epsilon_pL_s=(\epsilon_p/\epsilon_{\rm sn})L_{\rm sn}$ via Equation~\ref{shockpower}). Nevertheless, our result, say $\epsilon_p\lesssim0.1$ more conservatively, clearly suggests that we have reached the interesting parameter space for the purpose of probing the CR ion acceleration in embryonic SNRs embedded in a high-density material.  

The constraints on $\epsilon_p$ are quite insensitive to $s_p$, because the predicted GeV gamma-ray spectra are similar for different values of $s_p$ (see Figure~7 below). Based on the results shown in Figure~6 from left to right, we obtain $\epsilon_p\leq0.047~(\epsilon_{\rm sn}/0.25)$ (for $s_p=2.0$), $\epsilon_p\leq0.052~(\epsilon_{\rm sn}/0.25)$ (for $s_p=2.2$), and $\epsilon_p\leq0.073~(\epsilon_{\rm sn}/0.25)$ (for $s_p=2.4$), respectively. In any case, given that there are uncertainties in the source parameters used in our theoretical model, we conclude that the CR ion energy fraction in SN 2010jl is constrained to be less than $\sim10$\%. Note that Figure~1 corresponds to the case for $s_p=2.2$.

\begin{table*}[bt]
\begin{center}
\caption{Source parameters of SN 2010jl, which are used in the simplified model. The CSM size is assumed to be $\Delta R_s\approx R_s$. 
The derived shock power corresponds to $\epsilon_{\rm sn}=0.25$.  
}

\begin{tabular}{|c||c|c|c||c|c||c||c|c|}
\hline Name & $n_N$ [cm] & $R_s$ [cm] & $V_s$ [${\rm cm}~{\rm s}^{-1}$] & $L_{\rm sn}$ [erg/s] & ${\mathcal T}_{\rm sn}$ [K] & $\epsilon_{B}$ & $\epsilon_{p}$ & $\epsilon_{e}$\\
\hline SN 2010jl & $1.8\times{10}^{9}$ & $1\times{10}^{16}$ & $4\times{10}^8$ & $2.7\times{10}^{43}$ & $9000$ & $1.5\times{10}^{-2}$ & 0.05 & $5\times{10}^{-4}$\\
\hline
\end{tabular}

\end{center}
\end{table*}

\begin{table*}[bt]
\begin{center}
\caption{Source parameters of SN 2014C, which are used in the simplified model. Motivated by observations~\citep{Margutti:2016wyh}, we assume an CSM shell with ${\Delta R}_s=1.0\times{10}^{16}$~cm. 
}

\begin{tabular}{|c||c|c|c||c|c||c||c|c|}
\hline Name & $n_N$ [cm] & $R_s$ [cm] &  $V_s$ [${\rm cm}~{\rm s}^{-1}$] & $L_{\rm sn}$ [erg/s] & ${\mathcal T}_{\rm sn}$ [K] & $\epsilon_{B}$ & $\epsilon_{p}$ & $\epsilon_{e}$\\
\hline SN 2014C & $3.5\times{10}^{6}$ & $6.4\times{10}^{16}$ & $4\times{10}^{8}$ & $7\times{10}^{40}$ & $600$ & $1.5\times{10}^{-2}$ & 0.05 & $2\times{10}^{-4}$\\
\hline
\end{tabular}

\end{center}
\end{table*}

\begin{figure}[t]
\includegraphics[width=\linewidth]{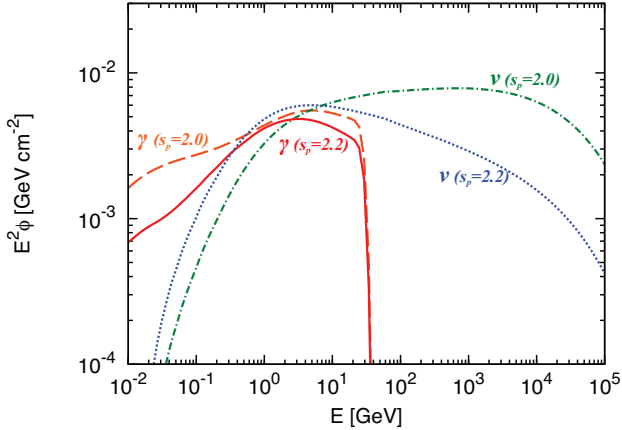}
\caption{Gamma-ray and (all-flavor) neutrino spectra expected for SN 2010jl. This panel shows results for different CR spectral indices, and GeV gamma-ray fluences are insensitive to $s_{p}$.
The simplified model is used. Both electromagnetic cascades in the radiation zone and attenuation in the pre-shock CSM are taken into account. The energy fraction carried by CRs is $\epsilon_p=0.05$.
}
\end{figure}

\begin{figure}[t]
\includegraphics[width=\linewidth]{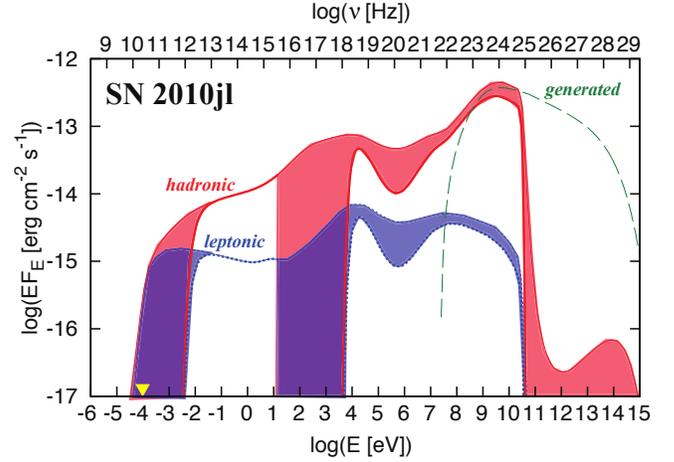}
\caption{Broadband non-thermal spectra from SN IIn 2010jl at $t=300$~d. The simplified model is used. The upper curves of the shaded regions indicate fluxes from the radiation zone, wheres the lower curves corresponds to the cases where attenuation in the pre-shock CSM is implemented. The CR parameters are $\epsilon_p=0.05$ and $s_p=2.2$. The source distance is $d=49$~Mpc. 
Radio upper limit (indicated by the triangle) taken from \cite{Chandra:2015jsa,Chandra:2017aev}.
}
\end{figure}

\begin{figure}
\includegraphics[width=\linewidth]{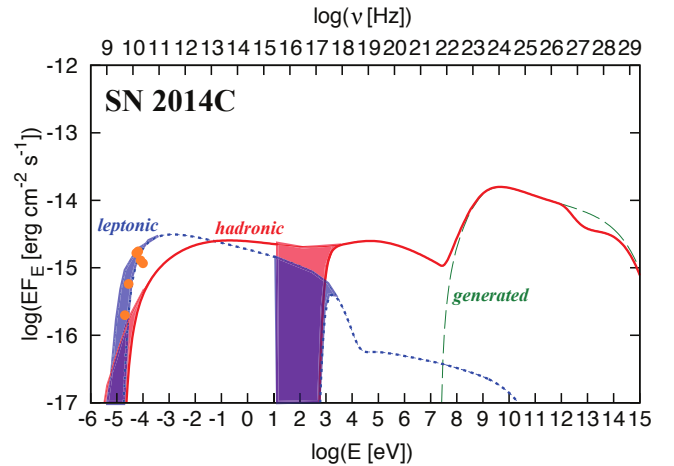}
\caption{Broadband non-thermal spectra from strongly interacting SN Ib 2014C at $t=396$~d. The simplified model is used. The CR parameters are $\epsilon_p=0.05$ and $s_p=2.2$. The source distance is $d=15.1$~Mpc. 
Radio data (indicated by the filled circles) taken from \cite{Margutti:2016wyh}.
}
\end{figure}

\section{Simplified model}
In the previous sections, we described the time-dependent model to calculate non-thermal emissions from interacting SNe, and considered SN 2010jl as one of the applications. As shown in \cite{Murase:2017pfe}, such a time-dependent model is essential, because the detectability of high-energy neutrinos and gamma rays depends on the signal-to-background ratio for such {\it long-duration transients}. 
However, detailed SN data in the optical band and/or at other wavelengths may not always be available. Then, a simpler model can still be useful in such cases where the quality of the observational data is rather limited. 
In the following, we consider a simplified version of the model (which is essentially a single-zone non-thermal radiation model with external thermal radiation fields), for a given set of the principal parameters ($n_N$, $R_s$, and $V_s$) that represent typical values during the time interval of interest and may be obtained observationally for a specific characteristic period. The setup is similar to that in \cite{Murase:2010cu}, but we present more detailed calculations on electromagnetic cascades as well as attenuation during the photon propagation in the pre-shock CSM. 

First, we apply the simplified model to SN 2010jl as an example. The model parameters are listed in Table~2, which are based on the observations at $t\sim300$~d~\citep{Ofek:2013afa,Fransson:2013qya}. 
The neutrino and gamma-ray spectra are shown in Figure~7. One clearly sees the dependence of neutrino and gamma-ray fluences on $s_{p}$ from this panel. Very importantly, results on GeV gamma-ray fluences are insensitive to $s_{p}$.  With the same $\epsilon_p$, GeV neutrino and gamma-ray fluences are higher for larger values of $s_p$, but electromagnetic cascades are more important for harder indices, which compensates the fluence difference. The component of CR-induced cascades is clearly evident below the kinematic break of $\pi^0$ decay at $E_\gamma\approx m_{\pi^0}c^2/2\simeq67.5$~MeV.
The results agree with those of the time-dependent model (compare Figure~7 to Figure~1). This implies that the simplified model is valid enough for the purpose of understanding physical properties of the emission. 
However, in general, this depends on parameters, and the time-dependent model is always better given that sufficient data are available.

In Figure~8, we also show broadband non-thermal spectra for both hadronic and leptonic components.
Pionic gamma rays produced via $\pi^0\rightarrow\gamma\gamma$ are denoted by ``generated $\gamma$''. The spectra from the radiation zone are also depicted by the upper curves of the shaded regions, where the attenuation in the pre-shock CSM (i.e., the shock upstream) is not include yet. We find that low-frequency radio emission is modified compared to simple predictions of synchrotron emission from relativistic electrons with a simple power-law injection spectrum. This is because the Coulomb cooling of electrons and positrons becomes relevant in high-density environments, making the spectrum harder by distorting the lepton distribution and suppressing the resulting synchrotron emission.  
In addition, we expect non-thermal X-ray emission because relativistic electrons and positrons up-scatter SN photons to an energy of $\sim\gamma_h^23k{\mathcal T}_{\rm sn}\sim12~{\rm keV}~{\mathcal T}_{\rm sn,4}$, where $\gamma_h\approx68$. 
In reality, we expect strong photoelectric absorption and subsequent thermalization in the pre-shock CSM, so soft X rays are significantly suppressed in the SN 2010jl-like cases. Also, the non-thermal X-ray component can readily be overwhelmed by the thermal bremsstrahlung component. We also consider the leptonic scenario, in which the emission would be weaker for $\epsilon_e/\epsilon_p\sim{10}^{-3}-{10}^{-2}$. The radio emission is significantly suppressed by the synchrotron self-absorption as well as the free-free absorption in the pre-shock CSM.
Note that our model predicts the existence of {\it both} hadronic and leptonic components. However, for Type IIn SNe like SN 2010jl, the hadronic component is expected to be dominant over the entire energy range. Thus, the model prediction consisting of the sum of the two components is essentially the same as that of the hadronic scenario. The predicted radio flux is consistent with upper limits at $\sim200-300$~d, provided by \cite{Chandra:2015jsa,Chandra:2017aev}. See Figure~8, where the upper limit at 22.5~GHz, measured at 204~d after the explosion, is shown. Note that it also supports that the radio signals should be absorbed in the CSM. The model fluxes are even more strongly suppressed at lower frequencies, and the radio data at 22.5~GHz give the most relevant constraint on the model.

Second, we apply our code to SN 2014C that occurred at $d=15.1$~Mpc. This SN was initially classified as an SN Ib, but in a year it started showing strong interactions with a dense CSM with $M_{\rm cs}\sim(1.0-1.5){\rm M}_{\odot}$. 
The model parameters listed in Table~3 are taken from the observational data at $t\sim400$~d~\citep{Milisavljevic:2015bli,Margutti:2016wyh}. 
In this case, the observations indicated that the CSM has a ``shell-like'' structure, so we consider $\Delta R_s< R_s$ and the CSM mass should be written as $M_{\rm cs}\approx\Delta\Omega R_s^2\Delta R_s m_Hn_N$. 
The forward shock is non-radiative. X-ray emission, which is consistent with bremsstrahlung emission, was observed by {\it Chandra} and {\it NuSTAR}~\citep{Margutti:2016wyh}. We here implement the X-ray luminosity, $L_X=5\times{10}^{40}~{\rm erg}~{\rm s}^{-1}$, and the X-ray temperature, $k{\mathcal T}_X=18$~keV, inferred from these observations. In any case, the results on photon spectra are not affected by target photons in the X-ray range. 
Thermal emission in the optical band already declined at such late times, and the long-lasting dust emission was observed in the infrared band. For simplicity, we use $L_{\rm sn}=7\times{10}^{40}~{\rm erg}~{\rm s}^{-1}$ and ${\mathcal T}_{\rm sn}=600$~K following \cite{Tinyanont:2016vbz}. 
The resulting spectra are shown in Figure~9. In this case, the gamma-ray emission would be too dim to detect because the CSM density is not high enough for the CR calorimetry to hold. The pionic gamma-ray component can be attenuated by the infrared photons only modestly in the 10~TeV range. Because of an electromagnetic cascade and the inverse-Compton scattering of the dust photons, the hadronic photon spectrum is flatter than the leptonic one. See also Model B in \cite{Murase:2010cu}.  
The radio signal in the 1-10~GHz range was detected~\citep{Margutti:2016wyh}, which seems best explained by synchrotron emission from primary electrons (i.e., the leptonic interpretation).
Note that the dominance of the leptonic component at sufficiently low frequencies is one of our predictions, as long as the CSM density is so large that the suppression is too strong~\citep[see also][]{Murase:2013rfa}. The Lorentz factor of secondary electrons and positrons is fixed at $\sim m_\pi/(4m_e)$, so the hadronic component has a corresponding low-energy break, as discussed in Section~2.  Then the leptonic component eventually takes over at low frequencies, as one can see in Figure~9.
Although we show the case of ${\mathcal T}_{\rm cs}^u=3\times{10}^{5}$~K for demonstration, the free-free absorption can be more significant for a lower temperature of ${\mathcal T}_{\rm cs}^u\sim{10}^{4}-10^{5}$~K. More detailed modeling that takes into account the CSM geometry would be necessary to robustly constrain $\epsilon_p$ and $\epsilon_e$. 
In any case, the model (with an CR ion energy fraction of $\epsilon_p\sim0.01-0.1$) predicts a hard spectrum of radio emission from secondary electrons and positrons, with a detectable flux of $F_\nu\sim1$~mJy at $\sim100-300$~GHz. Thus, {\it higher-frequency} observations are essential to critically test the existence of the hadronic component.

\begin{figure}
\includegraphics[width=\linewidth]{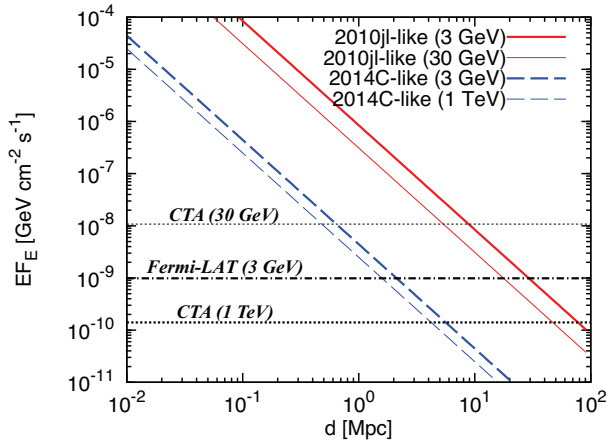}
\caption{The discovery horizon of hadronic gamma rays from interacting SNe with $\epsilon_p=0.1$. For theory lines, the simplified model is considered for both SN 2010jl-like and SN 2014C-like emissions. Differential sensitivities of {\it Fermi}-LAT (for an observation time of 1~yr based on the survey-mode Pass 8 sensitivity at the north Galactic pole) and CTA (for an observation time of 50~hr) are used~\citep{Consortium:2010bc}. 
}
\end{figure}

\section{Gamma-Ray Detectability of Nearby Interacting Supernovae}
Can we detect high-energy gamma-ray emission from nearby extragalactic SNe? For ordinary Type II SNe, the {\it Fermi}-LAT detection is possible up to a few~Mpc~\citep{Murase:2017pfe}, for CSM environments suggested by SN 2013fs. Type IIn SNe with a much higher CSM density, which this work focuses on, are more promising targets for the purpose of detecting extragalactic SNe. In order to address this question, using SN 2010jl and SN 2014C as gamma-ray spectrum templates (see Section~4), we compare the theoretical fluxes to differential sensitivities of {\it Fermi} LAT and Cherenkov Telescope Array (CTA). See Figure~10. For the SN 2010jl-like cases, GeV gamma-ray signal can be seen by {\it Fermi} up to $\sim30-50$~Mpc, whereas detections with imaging atmospheric Cherenkov telescopes may be more difficult especially in the TeV range because of the severe gamma-ray attenuation due to the two-photon annihilation process. 
These features are qualitatively consistent with the previous findings by \cite{Murase:2010cu}.
Note that we have used differential sensitivities, which give conservative estimates on the detectability. Indeed, the discovery horizons indicated in Figure~10 are somewhat worse than $\sim50$~Mpc suggested from the dedicated analysis presented in Section~3. 
For the SN 2014C-like cases, TeV gamma-ray attenuation in the pre-shock CSM is irrelevant. Therefore, observations with ground Cherenkov telescopes will be more powerful, and CTA can detect the interacting SNe up to $\sim5-8$~Mpc.
The rate of Type IIn is about $\sim7-9$\% of all the core-collapse SNe~\citep[e.g.,][]{Smith:2010vz,Li:2010kc}, and the occurrence rate of nearby Type IIn SNe within 30~Mpc is $\sim1~{\rm yr}^{-1}$. The rate of Type IIn SNe within 10~Mpc is thus $\sim0.03~{\rm yr}^{-1}$, but this value is conservative because the SN rate density within 10 Mpc is higher than the global one~\citep{Horiuchi:2011zz}. Because high-energy neutrinos from Type IIn are detectable by IceCube up to $\sim10$~Mpc~\citep{Murase:2010cu,Petropoulou:2017ymv}, interacting SNe should be regarded as promising sources for multi-messenger observations. 

\section{Summary}
Our results are summarized as follows.\\
(a) We numerically calculated broadband spectra of CR-induced non-thermal emission from interacting SNe, taking into account various processes such as electromagnetic cascades in the post-shock CSM as well as the attenuation in the pre-shock CSM. The electromagnetic cascade is unavoidable, although its detailed effect was not included in the previous literature~\citep{Murase:2010cu,Zirakashvili:2015mua,Petropoulou:2016zar}. We found that GeV gamma-ray spectra are insensitive to $s_p$ thanks to the cascade, and the attenuation effect can reduce the flux only modestly, which ensures that gamma rays can be used as a probe of shock interactions in dense environments that are difficult to directly observe in visible light, X rays, and radio waves. The phenomenological prescription we presented in this work enables us to robustly predict high-energy non-thermal signatures only with a few free parameters, given that the SN dynamics is determined by optical and X-ray observations. 
In the early stages of interactions, the deceleration of a fast velocity component of the SN ejecta dominates the dissipation. This enhances gamma-ray and neutrino fluxes by about one order of magnitude~\citep{Murase:2017pfe}, which can be taken into account our formalism. 
\\
(b) We applied our phenomenological time-dependent model to SN IIn 2010jl, by which for the first time we derived the gamma-ray constraint on the CR energy fraction. This cannot be done without the detailed modeling of high-energy emission. In addition, we re-analyzed the {\it Fermi}-LAT data taking advantage of the new Pass 8 data release, and updated the gamma-ray limits themselves. With both theoretical and observational improvements, we obtained the new constraint on the CR energy fraction, $\epsilon_p\lesssim0.05-0.1$. Within uncertainties, our results are consistent with the DSA theory of CRs as well as the observations of Galactic SNRs such as Cas A, but can be regarded as intriguing constraints on the CR ion acceleration in early SNe in dense environments.\\
(c) We considered both hadronic and leptonic components with the simplified model, and discussed the detectability of gamma-ray emission from nearby interacting SNe. High-energy gamma-ray and neutrino signals from Type IIn SNe may be observed in the near future.\\
(d) High-frequency radio emission from secondary electrons and positrons is a promising signature of CR ion acceleration in interacting SNe, although the CSM geometry is important for the detectability. Although detailed modeling is left for future work, our model predictions are consistent with the available radio data of SN IIn 2010jl and strongly interacting SN Ib 2014C. 

If the DSA theory is correct, it is natural to expect efficient acceleration of CR ions even in early stages of SNe. Particle acceleration at high-density environments has been suggested by observations of gamma rays from novae~\citep{Ackermann:2014vfa}. While the SNe and novae could still share some features, the nova shock is usually slower and other physical conditions may differ from the SN case, so both observations are relevant as independent information. 
For novae, the recent concurrent observation of optical and gamma-ray emission indicates $\epsilon_p\sim1$\% assuming that the optical emission is powered by the shock~\citep{Li:2017crr}. Although the gamma-ray attenuation is more important at lower-velocity shocks in general~\citep[][]{Murase:2013kda}, such a low efficiency could be an outcome of the magnetic field perpendicular to the shock normal. However, for interacting SNe, the magnetic field geometry is highly uncertain. The shock is more powerful, and CSM eruptions would be transient, also involving shocks and turbulences. Thus, constraints or detections will help us understand the physics of the DSA as well as pre-SN mass-loss mechanisms. In addition, these are important to address the origin of CRs around or beyond the knee energy and figure out their contribution to the diffuse neutrino flux observed in IceCube~\citep{Sveshnikova:2003sa,Murase:2013kda,Zirakashvili:2015mua,Petropoulou:2017ymv}. 
Astrophysical sources producing neutrinos via inelastic $pp$ collisions, if they significantly contribute to the diffuse neutrino flux, cannot avoid constraints from the isotropic diffuse gamma-ray background in the GeV-TeV range~\citep{Murase:2013rfa}. Our results indicate SNe IIn like SN 2010jl can serve as hidden CR accelerators~\citep{Murase:2015xka} just after the photon breakout, so they can alleviate the gamma-ray limits even though only a moderate fraction of the GeV gamma rays can be attenuated. On the other hand, SNe like SN 2014C do not suffer from the gamma-ray attenuation, so the diffuse gamma-ray background constraints can be important.    
High-energy emission from ordinary SNe such as Type II-P, II-L, and IIb SNe has also been predicted to be detectable for nearby SNe~\citep{Murase:2017pfe}, and our method is applicable to these objects.  
Future multi-messenger observations will give us important clues to this long-standing enigma of particle astrophysics. 

\begin{acknowledgements}
Kohta Murase thanks Poonam Chandra, Boaz Katz, Eran Ofek, and Todd Thompson for useful discussions. We also thank internal referees in the {\it Fermi}-LAT Collaboration, especially Regina Caputo, Seth Digel, Nicola Omodei, and Nicolas Renault-Tinacci for reviewing the manuscript. 
The work of Kohta Murase is supported by NSF Grant No. PHY-1620777 and the Alfred P. Sloan Foundation. He also acknowledges the Hubble Fellowship through the NASA and STScI during the early stages of this work. 
A.F. was supported by the Initiative and Networking Fund of the Helmholtz Association. 
Keiichi Maeda acknowledges support by JSPS KAKENHI Grant (18H04585, 18H05223, 17H02864).
J.F.B. is supported by NSF Grant No. PHY-1714479.
The preliminary estimates and results were presented at the PCTS workshop at Princeton University in 2015, AMON workshop at Penn State University in 2015, and the supernova workshop at ISSI in 2016. 

The \textit{Fermi}-LAT Collaboration acknowledges generous ongoing support
from a number of agencies and institutes that have supported both the
development and the operation of the LAT as well as scientific data analysis.
These include the National Aeronautics and Space Administration and the
Department of Energy in the United States, the Commissariat \`a l'Energie Atomique
and the Centre National de la Recherche Scientifique Institut National de Physique
Nucl\'eaire et de Physique des Particules in France, the Agenzia Spaziale Italiana
and the Istituto Nazionale di Fisica Nucleare in Italy, the Ministry of Education,
Culture, Sports, Science and Technology (MEXT), High Energy Accelerator Research
Organization (KEK) and Japan Aerospace Exploration Agency (JAXA) in Japan, and
the K.~A.~Wallenberg Foundation, the Swedish Research Council and the
Swedish National Space Board in Sweden.
Additional support for science analysis during the operations phase is gratefully
acknowledged from the Istituto Nazionale di Astrofisica in Italy and the Centre
National d'\'Etudes Spatiales in France. This work performed in part under DOE
Contract DE-AC02-76SF00515.
\end{acknowledgements}

\bibliographystyle{apj_8}
\bibliography{kmurase.bib}

\end{document}